\documentclass[pra,twocolumn,
 superscriptaddress,amsmath,amssymb,nofootinbib]{revtex4-2}
\usepackage{amsfonts}
\usepackage{amsmath}
\usepackage{bbm}
\usepackage{amssymb}
\usepackage{graphicx}%
\usepackage{footmisc}
\usepackage{bm}
\usepackage{braket}
\usepackage{CJK}
\usepackage{hyperref}
\usepackage{cleveref}
\usepackage{longtable,booktabs}
\usepackage[all,arc,knot,matrix]{xy}

\crefname{equation}{Eq.}{Eqs.}

\newcommand{\norm}[1]{\lVert#1\rVert}
\newcommand{\Tr}{\text{Tr}}

\begin{document}
\begin{CJK*}{UTF8}{gbsn}
\title{Multiple entropy production for multitime quantum processes}

\author{Zhiqiang Huang (黄志强)}
\email{hzq@wipm.ac.cn}
\affiliation{State Key Laboratory of Magnetic Resonance and Atomic and Molecular Physics, Innovation Academy for Precision Measurement Science and Technology, Chinese Academy of Sciences, Wuhan 430071, China}

\date{\today}

\begin{abstract}
    Entropy production and the detailed fluctuation theorem are of fundamental importance for thermodynamic processes. In this paper, we study the multiple entropy production for multitime quantum processes in a unified framework. For closed quantum systems and Markovian open quantum systems, the given entropy productions all satisfy the detailed fluctuation relation. This also shows that the entropy production rate under these processes is non-negative.  For non-Markovian open quantum systems, the memory effect can lead to a negative entropy production rate. Thus, in general, the entropy production of the marginal distribution does not satisfy the detailed FT relation. Our framework can be applied to a wide range of physical systems and dynamics. It provides a systematic tool for studying entropy production and its rate under arbitrary quantum processes.
\end{abstract}


\maketitle

\section{Introduction}\label{INTRO}
The fluctuation theorem can give a generalisation of the second law of thermodynamics and imply the Green-Kubo relations. It applies to fluctuations far from equilibrium and is of fundamental importance to non-equilibrium statistical mechanics. Roughly speaking, the fluctuation theorems (FTs) are closely related to time-reversal symmetry and the relations between the probabilities of forward and backward processes. For isolated quantum systems, the forward and backward processes can be described by unitary evolution. And there is always a widely held detailed FT \cite{EHM09}. For open quantum systems, one can assume that the entire system-environment combination is a large closed system and make use of the detailed FT of the closed system. Within this framework, the state of the environment must be detectable. If this is not the case, the backward mapping cannot fully recover the system state due to the lack of the environment information. One approach is to use the Petz recovery map as the backward processes and to establish the relation between the quantum channel and its Petz recovery map \cite{KK19}. These two approaches give the same entropy production when dealing with the maps with global fixed points \cite{LP21}.

The FTs focus mainly on entropy production. If an entropy production satisfies the detailed, we call it a fluctuation quantity here.
In the single-shot scenario, the entropy production depends on two-point measurements. For the multitime processes, the intermediate measurements may affect the system state and subsequent evolution, so the entropy production may depend on multi-point measurements. The process in presence of feedback control is a typical multitime process. And the corresponding FTs need to be modified to take into account the information gained from the measurement \cite{HV10,LRJ12,CS18}. With the multipoint measurements, it becomes more natural to study the entropy production rate. An important observation is that the non-negativity of the entropy production does not guarantee the non-negativity of the entropy production rate. The entropy production rate is determined by the entropy production relation between the $k+1$ step processes and the $k$ step processes. Since non-negative average entropy production is a natural consequence of the FT, studying the entropy production and the detailed FT for multitime processes can help to understand the relationship between the sign of the entropy production rate and the occurrence of non-Markovian effects. 

It should be noted that in the single-shot scenario, the entropy production rate can also be discussed by comparing the average entropy production for different evolutionary times. However, as we will explain later, evolution with different evolution times can be described by the same evolution process, but the measurement process is completely different. Therefore, entropy production at different evolutionary times does not correspond to the same overall process and cannot be described by the same joint distribution.

Usually the FT is directly related to the actual observation, and the multi-point measurements can give a joint probability distribution that can reflect the multi-time properties of the system. However, in the quantum regime, the measurements are generally invasive: the measurements are invasive not only to the system itself, but also to the subsequent dynamics of open system. On the one hand, since the measurement is invasive to the state, the measurement contributes directly to the entropy production. The cost of quantum measurement in a thermodynamic process are addressed by \cite{DPZ16,D21}. So \cite{DPZ16} tries to use a single measurement and gets a Jarzynski-like equality. Since there is only a single point measurement, the properties it gives must also depend only on a single point. There will be neither the concept of the backward processes, nor a fluctuation-dissipation theorem related to the properties of two-point measurements. On the other hand, because the measurement is invasive to the subsequent dynamics of open system, it must have other indirect effects on the entropy production of multi-time processes. Therefore, in this paper, we will consider the combined effects of evolution and measurement on the entropy production and the FT. 

The quantum dynamical semigroups are standard Markovian quantum processes. Their entropy production and detailed FT have been studied by \cite{JPW14,BB20}. For non-Markovian quantum processes, the memory effect makes the evolution much more complex. The process tensors \cite{PT,PTR2} are powerful operational tools for studying various temporally extended properties of general quantum processes. Using these tools, \cite{S19} set up a framework for quantum stochastic thermodynamics and discussed the entropy production of the Markovian processes. In our previous work \cite{H22}, we used an equivalent form of process tensors to obtain the FTs for non-Markovian processes. In this work, we continue to use this form and consider the marginal distribution of the multitime quantum processes. The detailed FT of the joint probability does not guarantee that the marginal distribution also satisfies these relations. If these relations are indeed satisfied, then there can be several compatible fluctuation quantities in the same processes. As we will show later, the existence of multiple compatible fluctuations is directly related to the issue of the entropy production rate. 

Due to the invasiveness of the measurements, the marginal distributions do not correspond to derived processes, in which some measurements are not performed. Since the memory effect can lead to a negative entropy production rate. Thus, in general, the entropy production of the marginal distribution does not satisfy the detailed FT relation.  Only if the Kolmogorov consistency condition is satisfied, then not performing a measurement is the same as averaging over its probabilities \cite{Plenio20}. And the entropy production of the corresponding marginal distributions will be the same as that of the derived processes. The entropy production of the derived processes should satisfy the detailed FT relation, so the entropy production of the marginal distributions should also do so. Another interesting relationship between the Kolmogorov condition and quantum thermodynamics is work extraction \cite{MA18}. Work extraction itself is not a fluctuation quantity and the proof of \cite{MA18} has nothing to do with backward processes. So the relation between the sum of each intermediate amount entropy production and the total entropy production is still unclear. And we will discuss it briefly in this paper.

This paper is organised as follows. In \cref{S2}, we first briefly introduce the general framework of operator states and process states. Then, for closed quantum systems, Markovian open quantum systems and non-Markovian open quantum systems, we try to derive the entropy production and the detailed FT of the joint probability and marginal distributions. We show that the Kolmogorov condition will make the sum of each intermediate amount entropy production equal to the total entropy production in the closed system, but fails for other systems. We also show how multiple compatible fluctuations are related to the entropy production rate. In \cref{EX}, we discuss the average entropy production of a simple Jaynes-Cummings model. The section \ref{CD} concludes.

\section{Entropy production for multitime quantum processes}\label{S2}
In the operator-state formalism \cite{PCASA19,H22}, operators are treated as states. The inner product of these states is defined as
\begin{equation}
    (O_1|O_2)=\text{Tr}(O_1^\dag O_2).
\end{equation}
The operator vector space is orthonormalized as follows:
\begin{equation}
    (\Pi_{kl}|\Pi_{ij})=\delta_{ik}\delta_{jl},
\end{equation}
where $\Pi_{ij}=\ket{i}\bra{j}$. The completeness relation is
\begin{equation}
    \hat{I}=\sum_{ij}|\Pi_{ij})(\Pi_{ij}|.
\end{equation}
The evolution of the system can be generally described with quantum channel $\mathcal{N}$, which is a superoperator that maps a density matrix $|\rho)$ to another density matrix  $|\rho')=\mathcal{N}|\rho)$. The operator state $|\Phi^{AS})=\sum_{ij}|\Pi_{ij}^A\otimes\Pi_{ij}^S)$ is often used to link the input and output of the state. It differs from the maximally entangled state by only a normalization factor $N$. Hence, the state $|\Psi^{AS}_\mathcal{N})\equiv\mathcal{N}^S|\Phi^{AS})/N$ is nothing but the Choi state obtained from the Choi-Jami\l kowski isomorphism. We will use the Choi-state form of the process tensors which we call the process states. This form can help us separate measurement from evolution, and it is more convenient to use the results about the Choi-state of quantum channel. 

\begin{table*}[!ht]
    \centering
    \begin{tabular}{|l|l|l|l|}
    \hline
        Time Evolution & Unitary & Markovian & Non-Markovian \\ \hline
        Satisfy the FT relation  & $R$,$R_1$,$R_2$,\dots & $R$,$R_1$,$R_2$,\dots & $R$, Others to be determined \\ \hline
        Average relation & $\braket{R}=\braket{R_1}+ \braket{R_2}$ & $\braket{R}\leq\braket{R_1}+ \braket{R_2}$& Depends on conditions \\ \hline
        Subsequent entropy production & $\braket{R}\geq \braket{R_{sub}}$ & $\braket{R}\geq \braket{R_{sub}}$ & Depends on conditions \\ \hline
    \end{tabular}
    \caption{The average entropy production relation of unitary evolution requires the Kolmogorov consistency condition.  For Markovian evolution, even if the measurement is non-invasive, due to the irreversibility of evolution, there will generally be $\braket{R}>\braket{R_1}+ \braket{R_2}$. One can find an entropy production such that $R'_1$+$R_2$=$R$, but the fluctuation relation of $R'_1$ is established in another process (\ref{APR1}).
    }
\end{table*}

\subsection{Closed quantum system}\label{CQS}

\begin{figure*}[htb]
    \centering
    \includegraphics[width=0.8\textwidth]{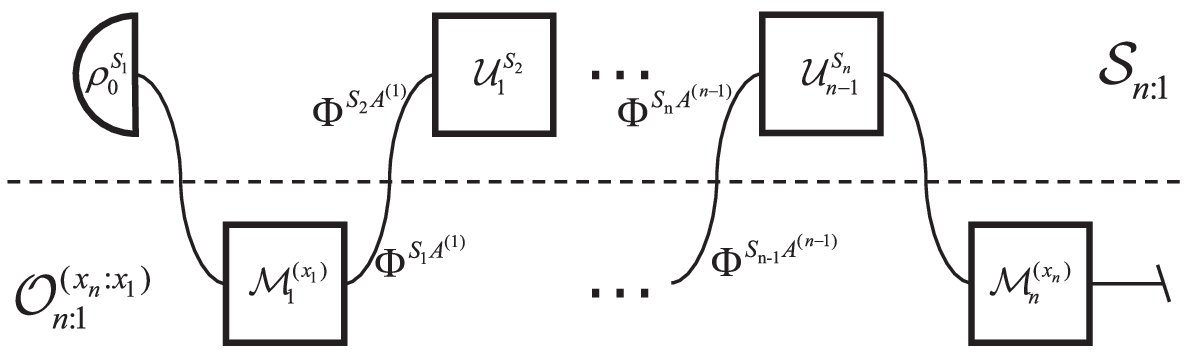}
    \includegraphics[width=0.8\textwidth]{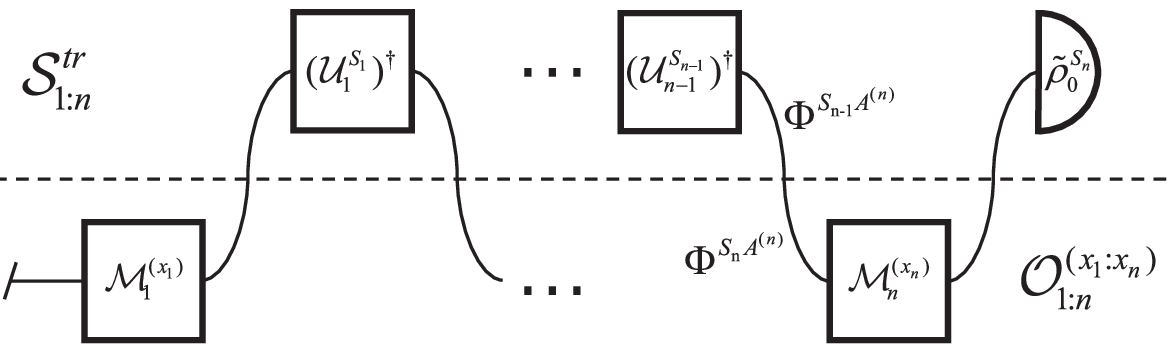}
    \caption{The forward process state $\mathcal{S}_{n:1}$ and the backward process state $\mathcal{S}^{tr}_{1:n}$ for  $n-1$ step unitary evolution. The inner product of $n$-point measurements operation $\mathcal{O}$ and process state gives joint probability distribution. The dashed line represents the separation of measurements and evolution.
    }
    \end{figure*}

In general, the time evolution of closed quantum systems is described by unitary operators acting on the system. Under $n-1$ step evolution, the forward process state is
\begin{equation}
    |\mathcal{S}_{n:1}):= \mathcal{U}^{S_{n} }_{n-1}\circ \ldots\circ \mathcal{U}^{S_2 }_1|\rho_0^{S_1}\otimes [\bigotimes_{j=2}^{n}  \Phi^{A^{(j-1)}S_{j}}]). 
\end{equation}
In such processes, it is natural to measure the state and use the measurement results as input for the next step. Therefore, we define the $n$-point measurements operation as
\begin{equation}\label{FNPO}
   (\mathcal{O}_{n:1}^{(x_n:x_1)}|:= (I^{S_n}\otimes[\bigotimes_{i=1}^{n-1}  \Phi^{S_{i}A^{(i)}}]|\mathcal{M}_n^{(x_n)}\otimes  \ldots \otimes \mathcal{M}_1^{(x_1)},
\end{equation}
where the operation $\mathcal{M}_i^{(x_i)}=|\Pi^{x_i})(\Pi^{x_i}|$ acts on $S_{i}$.
With these notations in hand, the joint probability for $n$-point measurements can be expressed as 
\begin{align}\label{JPFUP}
    \mathcal{P}_{n:1}(x_{n:1}|\mathcal{M}_{n:1})= (\mathcal{O}_{n:1}^{(x_n:x_1)}|\mathcal{S}_{n:1})\notag\\
    =(\Pi^{x_n}|\mathcal{U}_{n-1}|\Pi^{x_{n-1}})\ldots(\Pi^{x_2}|\mathcal{U}_1|\Pi^{x_1})(\Pi^{x_1}|\rho_0).
\end{align}
The unitary evolution is invertible with the time-reversed evolution. So the backward process state is
\begin{equation}
    |\mathcal{S}^{tr}_{1:n}):= (\mathcal{U}^{S_{n-1} }_{n-1})^\dagger\circ \ldots\circ (\mathcal{U}^{S_1 }_1)^\dagger|\tilde{\rho}_0^{S_n}\otimes [\bigotimes_{j=1}^{n-1}  \Phi^{A^{(j+1)}S_{j}}]), 
\end{equation}
where $\tilde{\rho}_0$ is the initial state of the backward process, and $\mathcal{U}^\dagger$ is the adjoint map. For single-shot evolution, the final state of the forward process is usually chosen as the initial state of the backward process. For open quantum systems, there are also some other choices \cite{LP21}. Here we chose 
\begin{equation}\label{BIS}
    |\tilde{\rho}_0)=\mathcal{U}_{n-1}\circ \ldots\circ \mathcal{U}_1|\rho_0),
\end{equation}
which is the final state of the forward process without any control operations. The backward $n$-point measurements operation can be defined as
\begin{equation}\label{BNPO}
   (\mathcal{O}_{1:n}^{(x_1:x_n)}|:= (I^{S_1}\otimes[\bigotimes_{i=1}^{n-1}  \Phi^{S_{i+1}A^{(i+1)}}]|\mathcal{M}_n^{(x_n)}\otimes  \ldots \otimes \mathcal{M}_1^{(x_1)}.
\end{equation}
The joint  probability for backward process can be expressed as 
\begin{align}\label{JPBUP}
    \mathcal{P}^{tr}_{1:n}(x_{1:n}|\mathcal{M}_{1:n})= (\mathcal{O}_{1:n}^{(x_1:x_n)}|\mathcal{S}^{tr}_{1:n})\notag\\
    =(\Pi^{x_1}|\mathcal{U}_{1}^{-1}|\Pi^{x_{2}})\ldots(\Pi^{x_{n-1}}|\mathcal{U}_{n-1}^{-1}|\Pi^{x_n})(\Pi^{x_n}|\tilde{\rho}_0).
\end{align}
It is easy to see that
\begin{equation}\label{FBUR}
    (\Pi^{x_j}|\mathcal{U}_{j-1}|\Pi^{x_{j-1}})=(\Pi^{x_{j-1}}|\mathcal{U}_{j-1}^\dagger|\Pi^{x_j}).
\end{equation}
The entropy production is defined as usual as the logarithm of the ratio of the forward and backward probabilities 
\begin{equation}\label{FTQ}
    R(x_{n:1}):=\ln \frac{\mathcal{P}_{n:1}(x_{n:1}|\mathcal{M}_{n:1})}{\mathcal{P}^{tr}_{1:n}(x_{1:n}|\mathcal{M}_{1:n})}.
\end{equation}
It follows from \cref{JPFUP,JPBUP,FBUR} that
\begin{equation}\label{UPEPTT}
    R(x_{n:1})=R(x_n,x_1)=\ln \frac{(\Pi^{x_1}|\rho_0)}{(\Pi^{x_n}|\tilde{\rho}_0)},
\end{equation}
which depends only on local measurements of the initial and final states. The distribution of entropy production is given by
\begin{align}
    p(R)=\sum_{x_{1:n}}\mathcal{P}_{n:1}(x_{n:1}|\mathcal{M}_{n:1})\delta(R-R(x_n,x_1)), \\
    p^{tr}(R)=\sum_{x_{1:n}}\mathcal{P}^{tr}_{1:n}(x_{1:n}|\mathcal{M}_{1:n})\delta(R+R(x_n,x_1)).
\end{align}
It then follows that 
\begin{equation}\label{CDFT}
    p(R)=e^R p^{tr}(-R),
\end{equation}
which gives the detailed FT.

Now consider the marginal distribution of the forward and backward probability. Without loss of generality, we divide the overall process into two parts: the first $n-2$ steps and the $n-1$th step. Summing over outcomes of the last measurement $\mathcal{M}_n$, we obtain a marginal distribution of the forward probability
\begin{equation}
    \mathcal{P}_{n:1}(x_{n-1:1}|\mathcal{M}_{n:1}):=\sum_{x_n}\mathcal{P}_{n:1}(x_{n:1}|\mathcal{M}_{n:1}).
\end{equation}
The backward one can be similarly defined. With these two probabilities, we can define another entropy production
\begin{equation}
    R(x_{n-1:1}):=\ln \frac{\mathcal{P}_{n:1}(x_{n-1:1}|\mathcal{M}_{n:1})}{\mathcal{P}^{tr}_{1:n}(x_{1:n-1}|\mathcal{M}_{1:n})}.
\end{equation}
Similar to \cref{UPEPTT}, it is easy to show that
\begin{equation}\label{UETEP}
    R(x_{n-1:1})=R(x_{n-1},x_1)=\ln \frac{(\Pi^{x_1}|\rho_0)}{(\Pi^{x_{n-1}}|\tilde{\rho}_1)},
\end{equation}
where $|\tilde{\rho}_1)=\mathcal{U}_{n-1}^{-1}\circ \mathcal{M}_{n} |\tilde{\rho}_0)$. And $\mathcal{M}_{k}=\sum_{x_k}\mathcal{M}_{k}^{(x_k)}$ is a dephasing map. The entropy production $R(x_{n-1:1})$ depends only on local measurements of $\rho_0$ and $\tilde{\rho}_1$. The corresponding distribution of entropy production is given by
\begin{align}
    p(R_1)=\sum_{x_{1:n}}\mathcal{P}_{n:1}(x_{n:1}|\mathcal{M}_{n:1})\delta(R-R(x_{n-1},x_1)), \\
    p^{tr}(R_1)=\sum_{x_{1:n}}\mathcal{P}^{tr}_{1:n}(x_{1:n}|\mathcal{M}_{1:n})\delta(R+R(x_{n-1},x_1)).
\end{align}
The entropy production $R_1$ also satisfies the detailed FT
\begin{equation}
    p(R_1)=e^{R_1} p^{tr}(-R_1).
\end{equation}

If summing over outcomes of measurements $\mathcal{M}_{n-2:1}$, we obtain another marginal distribution of the forward probability
\begin{equation}
    \mathcal{P}_{n:1}(x_{n:n-1}|\mathcal{M}_{n:1}):=\sum_{x_{n-2:1}}\mathcal{P}_{n:1}(x_{n:1}|\mathcal{M}_{n:1}).
\end{equation}
Similar to previous procedures, we can define
\begin{equation}
    R(x_{n:n-1}):=\ln \frac{  \mathcal{P}_{n:1}(x_{n:n-1}|\mathcal{M}_{n:1})}{\mathcal{P}^{tr}_{1:n}(x_{n-1:n}|\mathcal{M}_{1:n})}
\end{equation}
and show that
\begin{equation}
    R(x_{n:n-1})=\ln \frac{(\Pi^{x_{n-1}}|\rho_{n-1})}{(\Pi^{x_n}|\tilde{\rho}_0)},
\end{equation}
where
\begin{equation}\label{URHON}
    |\rho_{n-1})=\mathcal{U}_{n-2}\circ\mathcal{M}_{n-2}  \ldots \circ \mathcal{U}_1\circ \mathcal{M}_1 |\rho_0).
\end{equation}
The corresponding distribution of entropy production is given by
\begin{align}
    p(R_2)=\sum_{x_{1:n}}\mathcal{P}_{n:1}(x_{n:1}|\mathcal{M}_{n:1})\delta(R-R(x_{n},x_{n-1})), \\
    p^{tr}(R_2)=\sum_{x_{1:n}}\mathcal{P}^{tr}_{1:n}(x_{1:n}|\mathcal{M}_{1:n})\delta(R+R(x_{n},x_{n-1})).
\end{align}
The entropy production $R_2$ also satisfies the detailed FT. As stated above, the multitime quantum processes allow multiple entropy production terms. They all satisfy the detailed FT in a common multitime process. The previous procedures are actually applicable to all the marginal distributions. The corresponding entropy production also depends on local measurements.

If the joint probability $\mathcal{P}_{n:1}$ and $\mathcal{P}^{tr}_{1:n}$ satisfy the Kolmogorov consistency condition \cite{Plenio20}, then measuring but summing the measurements is equivalent to not measuring. In such cases, the probability distributions for all
subsets of times can be obtained by marginalization. And the detailed FT mentioned above are all equivalent to two-point measurement FT of some processes. In addition, when the Kolmogorov condition is met, there will be $  |\rho_{n-1})=\mathcal{U}_{n-2}\circ \ldots\circ \mathcal{U}_1|\rho_0)$ and $|\tilde{\rho}_1)=\mathcal{U}_{n-1}^{-1}|\tilde{\rho}_0)$. If choosing the initial state of the backward process as \cref{BIS}, then we have 
\begin{equation}\label{FRR}
    |\tilde{\rho}_1)=(\mathcal{U}_{n-1}^{-1}\circ\mathcal{U}_{n-1})\circ \ldots\circ \mathcal{U}_1|\rho_0)= |\rho_{n-1})
\end{equation}
and $(\Pi^{x_{n-1}}|\rho_{n-1})=(\Pi^{x_{n-1}}|\tilde{\rho}_1)$. Under these circumstances, the previously mentioned entropy production terms have the following relation
\begin{align}\label{FTQR}
    R(x_{n-1:1}) + R(x_{n:n-1})= \ln\frac{(\Pi^{x_{n-1}}|\rho_{n-1})(\Pi^{x_1}|\rho_0)}{(\Pi^{x_n}|\tilde{\rho}_0)(\Pi^{x_{n-1}}|\tilde{\rho}_1)}\notag \\ 
    = \ln\frac{(\Pi^{x_1}|\rho_0)}{(\Pi^{x_n}|\tilde{\rho}_0)}= R(x_{n:1}).
\end{align}
With this relation, one can easily see that $\braket{R_1}+\braket{R_2}=\braket{R}$, which is very similar to the relation of average work done shown in \cite{MA18}. However, the work itself is not the entropy production. The relation of average work has nothing to do with the backward processes. Therefore, the two relations are very different.

Combining \cref{FTQR} and the detailed FT of $R(x_{n:n-1})$, we have
\begin{equation}\label{FTQR2}
    \braket{e^{-[R(x_{n:1})- R(x_{n-1:1})]}}=1.
\end{equation}
The condition (\ref{FTQR}) is strong, but it is not a necessary condition for \cref{FTQR2}. In fact, if both $R$ and $R_1$ satisfy the detailed FT, then we have
\begin{align}\label{RBTPSP}
    \braket{e^{-[R(x_{n:1})- R(x_{n-1:1})]}}\notag\\
    =\sum_{x_{1:n}} \mathcal{P}_{n:1}(x_{n:1}|\mathcal{M}_{n:1})e^{-[R(x_{n:1})- R(x_{n-1:1})]}\notag \\
   =\sum_{x_{1:n}} \mathcal{P}^{tr}_{1:n}(x_{1:n}|\mathcal{M}_{1:n})e^{R(x_{n-1:1})}\notag \\
   =\sum_{x_{1:n-1}} \mathcal{P}^{tr}_{1:n}(x_{1:n-1}|\mathcal{M}_{1:n})e^{R(x_{n-1:1})}\notag \\
   =\sum_{x_{1:n-1}} \mathcal{P}_{n:1}(x_{n-1:1}|\mathcal{M}_{n:1}) =1.
\end{align}
Using Jensen's inequality $\braket{e^X}\geq e^{\braket{X}}$, \cref{FTQR2} implies
\begin{equation}\label{TAEPNLIAAEP}
     0\leq \braket{R(x_{n:1})-R(x_{n-1:1})}=\braket{R}-\braket{R_1},
\end{equation}
which means the total average entropy production is not less than intermediate average entropy production. This also implies that the entropy production rate at step $n-1$ is nonnegative. The proof (\ref{RBTPSP}) applies to all cases where the joint probability is well-defined.

Now let's calculate the average of entropy production. The aforementioned Kolmogorov consistency condition and \cref{BIS} are mainly used to give $(\Pi^{x_{n-1}}|\rho_{n-1})=(\Pi^{x_{n-1}}|\tilde{\rho}_1)$. Without these conditions, using \cref{JPFUP,UPEPTT,URHON}, the average entropy production $R$ can be expressed generally as
\begin{align}\label{EPRA}
    \braket{R}=\Tr[(\mathcal{M}_1 \rho_0) \ln (\mathcal{M}_1 \rho_0)]-\Tr[(\mathcal{M}_{n} \rho_{n}) \ln (\mathcal{M}_{n}\tilde{\rho}_0)] \notag\\
    =S(\mathcal{M}_{n} \rho_{n}||\mathcal{M}_{n}\tilde{\rho}_0) +S(\mathcal{M}_{n} \rho_{n})-S(\mathcal{M}_{1} \rho_{0})\notag\\
    =S(\mathcal{M}_{n} \rho_{n}||\mathcal{M}_{n}\tilde{\rho}_0) +\sum_{m=2}^{n}[S(\mathcal{M}_{m} \rho_{m})-S(\rho_m)],
\end{align}
where $S(\rho||\sigma):=\Tr[\rho(\ln \rho -\ln \sigma)]$ is the quantum relative entropy. In the third equal sign of \cref{EPRA} we exploit the fact that unitary transformations do not change entropy. For single-shot unitary evolution, if two-point measurements do not invade $\rho_0$, $\rho_n$ and $\tilde{\rho}_0$, then we can obtian the commonly used average entropy production \cite{EHM09}
\begin{equation}
    \braket{R}=S(\rho(t))||\tilde{\rho}_0).
\end{equation}
The average of the entropy production $R_1$ and $R_2$ is
\begin{align}\label{EPR12A}
    \braket{R_1}&=S(\mathcal{M}_{n-1} \rho_{n-1}||\mathcal{M}_{n-1}\tilde{\rho}_1)\notag\\
     &+S(\mathcal{M}_{n-1} \rho_{n-1})-S(\mathcal{M}_{1} \rho_{0}),\notag\\
     \braket{R_2}&=S(\mathcal{M}_{n} \rho_{n}||\mathcal{M}_{n}\tilde{\rho}_0)\notag\\
     &+S(\mathcal{M}_{n} \rho_{n})-S(\mathcal{M}_{n-1} \rho_{n-1}).
\end{align}
Using the non-negativity of relative entropy and the data processing inequality, it is easy to find that these average entropy productions are non-negative.
Combining \cref{EPRA,EPR12A}, it is easy to find that
\begin{equation}
    \braket{R_1}+ \braket{R_2}- \braket{R}=S(\mathcal{M}_{n-1} \rho_{n-1}||\mathcal{M}_{n-1}\tilde{\rho}_1)\geq 0.
\end{equation}
If the Kolmogorov consistency condition is not satisfied, then the sum of the segmental entropy production is generally greater than  the total average entropy production. 

\subsection{Markovian open quantum system}\label{OSMFT}
\begin{figure*}[htb]
    \centering
    \includegraphics[width=0.9\textwidth]{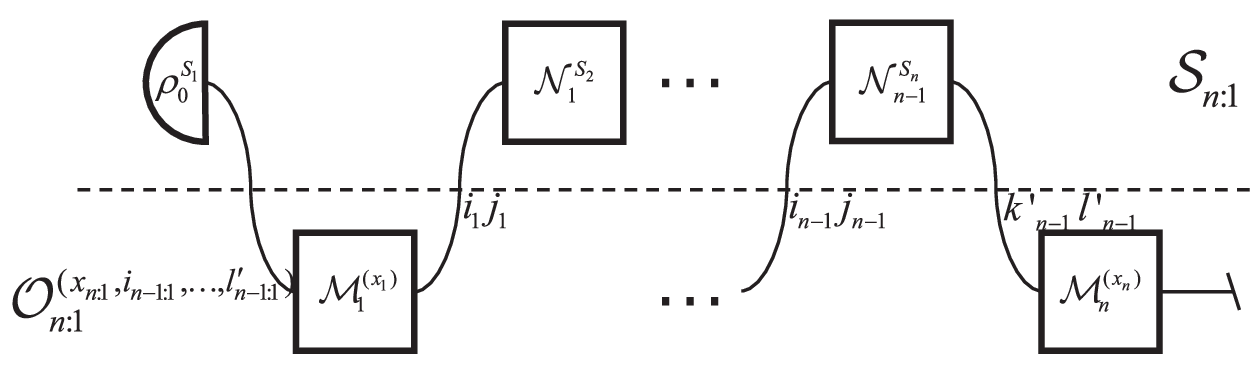}
    \includegraphics[width=0.9\textwidth]{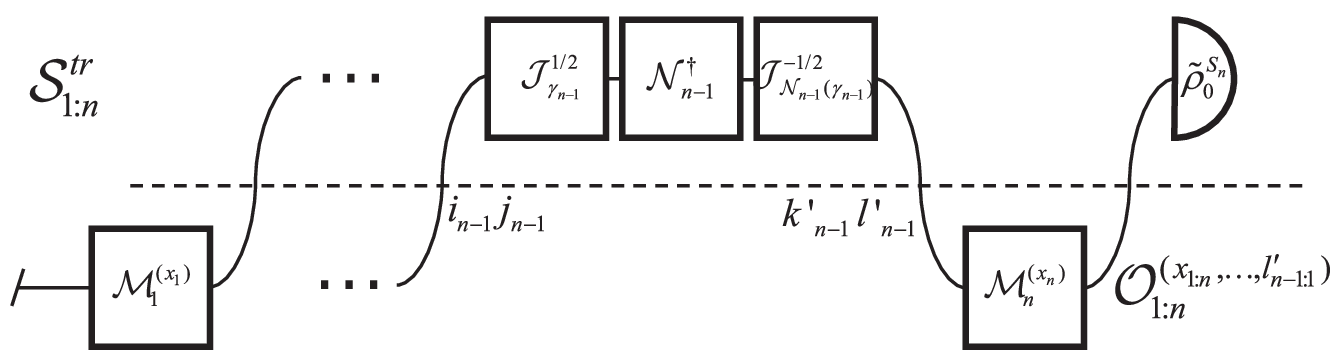}
    \caption{The forward and backward processes of $n-1$ step Markovian quantum evolution. The quasi-measurements make the link $\Phi^{AS}$ become $\Pi_{ij}^{A}\otimes\Pi_{ij}^{S}$ and  $\Pi_{k'l'}^{A}\otimes\Pi_{k'l'}^{S}$, which are abbreviated as $ij$ and $k'l'$ in the figure.}
    \end{figure*}
    The time evolution of the Markovian open quantum system can be described by a sequence of independent CPTP maps \cite{PRFPM18}. Under $n-1$ step evolution, the forward process state is
    \begin{equation}\label{NSMFPS}
    |\mathcal{S}_{n:1}):=\mathcal{N}^{S_{n} }_{n-1}\circ \ldots \circ\mathcal{N}^{S_{2} }_1|\rho_0^{S_1}\otimes [\bigotimes_{j=2}^{n}  \Phi^{A^{(j-1)}S_{j}}]). 
    \end{equation}
    With the measurements (\ref{FNPO}), the joint probability for $n$-point measurements can still be expressed as 
\begin{equation}\label{JPNPM}
    \mathcal{P}_{n:1}(x_{n:1}|\mathcal{M}_{n:1})= (\mathcal{O}_{n:1}^{(x_n:x_1)}|\mathcal{S}_{n:1}).
\end{equation}
    For open quantum systems, the lack of information about environment makes the evolution irreversible. It is common to use Petz recovery map as the backward map \cite{P86}.  For $n-1$ step evolution, the backward process state can be written as
    \begin{equation}\label{NSMBP}
    |\mathcal{S}^{tr}_{1:n}):=\mathcal{R}^{S_{n-1} }_{n-1}\circ \ldots \circ\mathcal{R}^{S_{1} }_1|\tilde{\rho}_0^{S_n}\otimes [\bigotimes_{j=1}^{n-1}  \Phi^{A^{(j+1)}S_{j}}]),
\end{equation}
where $\mathcal{R}_{m}:=\mathcal{J}^{1/2}_{\gamma_m} \circ\mathcal{N}_{m}^\dagger \circ \mathcal{J}^{-1/2}_{\mathcal{N}_{m}(\gamma_m)}$ is the Petz recovery map of $\mathcal{N}_{m}$, $\mathcal{J}^{\alpha}_{O}(\cdot):=O^\alpha (\cdot)O^{\alpha\dagger}$ is the rescaling map, and $\gamma_m$ is the reference state that can be freely chosen. The Petz recovery map is a CPTP map and fully recovers the reference state.  With the measurements (\ref{BNPO}), the joint  probability for backward process can be expressed as 
\begin{equation}\label{JPNPMB}
    \mathcal{P}_{n:1}(x_{n:1}|\mathcal{M}_{n:1})= (\mathcal{O}_{n:1}^{(x_n:x_1)}|\mathcal{S}_{n:1}).
\end{equation}
Similar to \cref{FBUR}, the adjoint map $\mathcal{N}^\dagger$ and $\mathcal{N}$ obey the following relation:
\begin{equation}\label{FBUOCR}
    (\Pi_{k'l'}|\mathcal{N}|\Pi_{ij})=(\Pi_{ij}|\mathcal{N}^\dagger|\Pi_{k'l'})^*.
\end{equation}
The rescaling map in Petz recovery will make the joint probability \cref{JPNPM,JPNPMB} generally unable to establish a relation similar to \cref{CDFT}. Only with the following operation can we obtain a detailed FT
\begin{align}\label{POVMNM}
    (\mathcal{O}_{n:1}^{(x_{n:1},i_{n-1:1},\ldots,l'_{n-1:1})}|:= (I^{S_n}\otimes[\bigotimes_{m=1}^{n-1} \Pi_{i_mj_m}^{S_{m}}\Pi_{i_mj_m}^{A^{(m)}} ]|\notag \\
    \mathcal{M}_n^{(x_n)}\mathcal{M}_{n}^{(k'_{n-1}l'_{n-1})}\otimes    \ldots \otimes  \mathcal{M}_2^{(x_2)} \mathcal{M}_2^{(k'_1l'_1)}\otimes  \mathcal{M}_1^{(x_1)},
 \end{align}
where $\mathcal{M}_{m+1}^{(k'_{m}l'_{m})}:=|\Pi_{k'_ml'_m}^{S_{m+1}})(\Pi_{k'_ml'_m}^{S_{m+1}}|$. The basis $\ket{i_m}$ is chosen such that it diagonalizes the reference state $\gamma_m$ and $\ket{k'_m}$ is chosen as the eigenbasis of $\mathcal{N}_{m}(\gamma_m)$. On this basis, we have $\mathcal{J}^{\alpha/2}_{\gamma_m}|\Pi_{i_mj_m})=Z_{ij}^{\gamma^{\alpha}}|\Pi_{i_mj_m}) $, where
 \begin{equation}
    Z_{ij}^{\gamma^{\alpha}} :=\norm{\mathcal{J}_{\gamma}^{\alpha/2}\Pi_{ij}}_2=\sqrt{ (\Pi_{i}|\gamma^{\alpha})(\gamma^{\alpha}|\Pi_{j})}.
 \end{equation}
 With the operation (\ref{POVMNM}), we obtain a quasiprobability distribution \cite{KK19}
\begin{align}\label{QSNM}
    \mathcal{P}_{n:1}(x_{n:1},i_{n-1:1},j_{n-1:1},k'_{n-1:1},l'_{n-1:1}|\mathcal{M}_{n:1}^{qs})\notag \\
    := (\mathcal{O}_{n:1}^{(x_{n:1},\ldots,l'_{n-1:1})}|\mathcal{S}_{n:1}).
\end{align}
This distribution is not positive, but it can be obtained from observable quantities \cite{KK19}. The operation (\ref{POVMNM}) can also be fully reconstructed as a linear combination of the $n$-point positive-operator valued measurements (POVMs) operation. Moreover, the joint probability can be directly derived from quasiprobability
\begin{equation}\label{POVMTOPVMMP}
    \sum_{i_{n-1:1},\ldots} \mathcal{P}_{n:1}(x_{n:1},\ldots,l'_{n-1:1}|\mathcal{M}_{n:1}^{qs})=\mathcal{P}_{n:1}(x_{n:1}|\mathcal{M}_{n:1}).
\end{equation}
The backward quasi-measurements operation can be defined as 
\begin{align}\label{BKPOVMNM}
   (\mathcal{O}_{1:n}^{(x_{1:n},\ldots,l'_{n-1:1})}|:= (I^{S_1}\otimes[\bigotimes_{m=1}^{n-1}  \Pi_{k'_ml'_m}^{S_{m+1}}\Pi_{k'_ml'_m}^{A^{(m+1)}}]|\notag \\
   \mathcal{M}_n^{(x_n)}\otimes\mathcal{M}_{n-1}^{(x_{n-1})}  \mathcal{M}_{n-1}^{(i_{n-1}j_{n-1})}\otimes \ldots \otimes \mathcal{M}_1^{(x_1)} \mathcal{M}_{1}^{(i_{1}j_{1})},
\end{align}
with which we obtain the quasiprobability distribution of backward processes
\begin{equation}\label{BQSNM}
    \mathcal{P}^{tr}_{1:n}(x_{1:n},\ldots,l'_{1:n-1}|\mathcal{M}_{1:n}^{qs}) := (\mathcal{O}_{1:n}^{(x_{n:1},\ldots,l'_{n-1:1})}|\mathcal{S}^{tr}_{1:n})^*.
\end{equation}
The joint probability of backward processes can also be directly derived from quasiprobability of backward processes like \cref{POVMTOPVMMP}.
Similar to (\ref{FTQ}), the entropy production of  which can be defined as 
\begin{equation}
    R(x_{n:1},\ldots,l'_{n-1:1}):=\ln \frac{\mathcal{P}_{n:1}(x_{n:1},\ldots,l'_{n-1:1}|\mathcal{M}_{n:1}^{qs})}{ \mathcal{P}^{tr}_{1:n}(x_{1:n},\ldots,l'_{1:n-1}|\mathcal{M}_{1:n}^{qs})}.
\end{equation}
With \cref{FBUOCR}, it is easy to show that
\begin{equation}
    (\Pi_{k'_ml'_m}|\mathcal{N}_m|\Pi_{i_mj_m})^*= (\Pi_{i_mj_m}|\mathcal{R}_m|\Pi_{k'_ml'_m})Z^{{\gamma}_m^{-1}}_{i_mj_m}Z^{\mathcal{N}_m(\gamma_m)}_{k'_ml'_m}.
\end{equation}
Combining this with \cref{QSNM,BQSNM}, we find that
\begin{align}\label{MFTQ}
    R(x_{n:1},\ldots,l'_{n-1:1})=R(x_n,x_1,\ldots,l'_{n-1:1})\notag \\
    =\ln \frac{(\Pi^{x_1}|\rho_0)}{(\Pi^{x_n}|\tilde{\rho}_0)}+\sum_{m=1}^{n-1}\ln(Z^{{\gamma}_m^{-1}}_{i_mj_m}Z^{\mathcal{N}_m(\gamma_m)}_{k'_ml'_m})
\end{align}
is independent of intermediate measurements $\{x_{n-1:2}\}$. The distribution of entropy production for forward processes is given by
\begin{align}
    p(R)=\sum_{x_{n:1},\ldots,l'_{n-1:1}}\mathcal{P}_{n:1}(x_{n:1},\ldots,l'_{n-1:1}|\mathcal{M}_{n:1}^{qs}) \notag \\
    \times\delta(R-R(x_n,x_1,\ldots,l'_{n-1:1})).
\end{align}
The backward one can be similarly defined. The detailed FT
\begin{equation}
    p(R)=e^{R} p^{tr}(-R)
\end{equation}
has been shown in Ref. \cite{H22}. 

Now follow the same procedure used in \cref{CQS}, the marginal distribution can be defined as
\begin{align}\label{QSIMP}
    \mathcal{P}_{n:1}(x_{n-1:1},i_{n-2:1},\ldots,l'_{n-2:1}|\mathcal{M}_{n:1}^{qs})\notag\\
    :=\sum_{x_n,i_{n-1},\ldots,l'_{n-1}}\mathcal{P}_{n:1}(x_{n:1},\ldots,l'_{n-1:1}|\mathcal{M}_{n:1}^{qs}).
\end{align}
The corresponding entropy production is
\begin{equation}
    R(x_{n-1:1},\ldots,l'_{n-2:1}):=\ln \frac{\mathcal{P}_{n:1}(x_{n-1:1},\ldots,l'_{n-2:1}|\mathcal{M}_{n:1}^{qs})}{\mathcal{P}^{tr}_{1:n}(x_{1:n-1},\ldots,l'_{1:n-2}|\mathcal{M}_{1:n}^{qs})},
\end{equation}
which is independent of intermediate measurements $\{x_{n-2:2}\}$:
\begin{align}\label{MFTQ1}
    R(x_{n-1:1},\ldots,l'_{n-2:1})=R(x_{n-1},x_1,\ldots,l'_{n-2:1})\notag \\
    =\ln \frac{(\Pi^{x_1}|\rho_0)}{(\Pi^{x_{n-1}}|\tilde{\rho}_1)}+\sum_{m=1}^{n-2}\ln(Z^{{\gamma}_m^{-1}}_{i_mj_m}Z^{\mathcal{N}_m(\gamma_m)}_{k'_ml'_m}),
\end{align}
where $|\tilde{\rho}_1)=\mathcal{R}_{n-1}\circ \mathcal{M}_n|\tilde{\rho}_0)$. The distributions of entropy production can be defined similarly to the previous one, and the detailed FT also holds. 

\begin{figure*}[htb]
    \centering
    \includegraphics[width=0.8\textwidth]{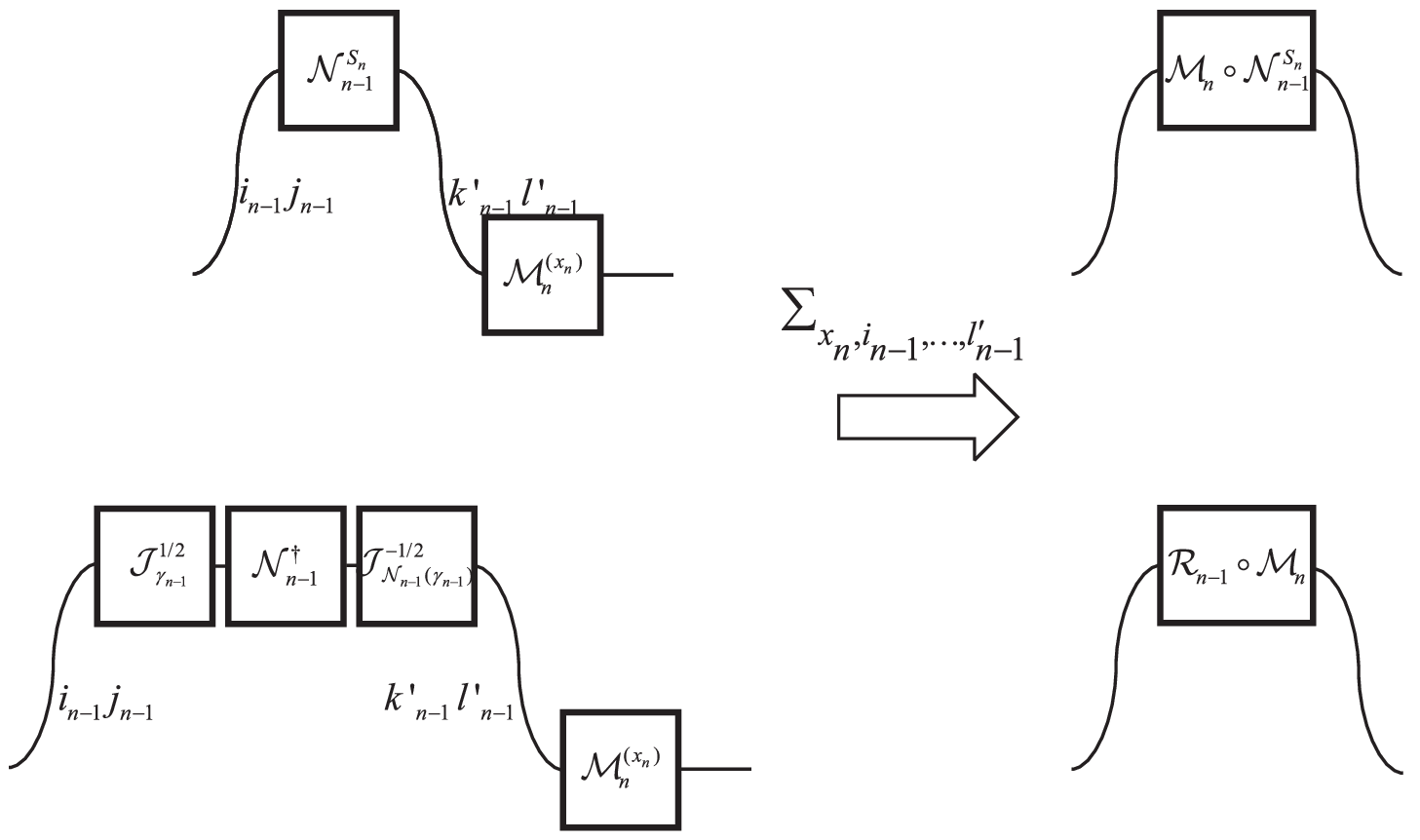}
    \caption{For Markovian processes, summing over outcomes of (quasi-)measurements gives a dephasing map.
    }
    \end{figure*}

After summing over outcomes of measurements $\mathcal{M}_{n-2:1}$, we obtain marginal distribution
\begin{align}
    \mathcal{P}_{n:1}(x_{n:n-1},i_{n-1},\ldots,l'_{n-1}|\mathcal{M}_{n:1}^{qs})\notag\\
    :=\sum_{x_{n-2:1},i_{n-2:1},\ldots,l'_{n-2:1}}\mathcal{P}_{n:1}(x_{n:1},\ldots,l'_{n-1:1}|\mathcal{M}_{n:1}^{qs}).
\end{align}
The corresponding entropy production is
\begin{equation}
    R(x_{n:n-1},\ldots,l'_{n-1}):=\ln \frac{ \mathcal{P}_{n:1}(x_{n:n-1},\ldots,l'_{n-1}|\mathcal{M}_{n:1}^{qs})}{\mathcal{P}^{tr}_{1:n}(x_{n-1:n},\ldots,l'_{n-1}|\mathcal{M}_{1:n}^{qs})}.
\end{equation}
It then follows that
\begin{align}\label{MFTQ2}
    R(x_{n:n-1},\ldots,l'_{n-1})=\ln \frac{(\Pi^{x_{n-1}}|\rho_{n-1})}{(\Pi^{x_n}|\tilde{\rho}_0)}\notag \\
    +\ln(Z^{{\gamma}_{n-1}^{-1}}_{i_{n-1}j_{n-1}}Z^{\mathcal{N}_{n-1}(\gamma_{n-1})}_{k'_{n-1}l'_{n-1}}),
\end{align}
where
\begin{equation}\label{MRHON}
    |\rho_{n-1})=\mathcal{N}_{n-2}\circ \mathcal{M}_{n-2} \ldots \mathcal{N}_1\circ \mathcal{M}_1|\rho_0).
\end{equation}
The distributions of entropy production can be defined similarly to the previous one, and the detailed FT also holds. 

Due to the irreversibility of open system evolution, there is no relation like \cref{FRR} for $\tilde{\rho}_1$ and $\rho_{n-1}$, even if the Kolmogorov condition is met. And the \cref{FTQR} no longer holds for the $\{R,R_1,R_2\}$ defined here. The backward process of $R_1$ is derived from $\mathcal{S}^{tr}_{1:n}$. If we use the following $n-2$ step backward processes instead
\begin{equation}\label{APR1}
    |\mathcal{S}^{tr}_{1:n-1}):=\mathcal{R}^{S_{n-2} }_{n-2}\circ \ldots \circ\mathcal{R}^{S_{1} }_1|\rho_{n-1}^{S_{n-1}}\otimes [\bigotimes_{j=1}^{n-2}  \Phi^{A^{(j+1)}S_{j}}]),
\end{equation}
then we can obtain quasiprobability distribution
\begin{align}
    \mathcal{P}^{tr}_{1:n-1}(x_{1:n-1},\ldots,l'_{1:n-2}|\mathcal{M}_{1:n-1}^{qs}) \notag\\
    := (\mathcal{O}_{1:n-1}^{(x_{n-1:1},\ldots,l'_{n-2:1})}|\mathcal{S}^{tr}_{1:n-1})^*
\end{align}
and another entropy production
\begin{align}
    R'(x_{n-1:1},\ldots,l'_{n-2:1})\notag\\
    :=\ln \frac{\mathcal{P}_{n:1}(x_{n-1:1},\ldots,l'_{n-2:1}|\mathcal{M}_{n:1}^{qs})}{\mathcal{P}^{tr}_{1:n-1}(x_{1:n-1},\ldots,l'_{1:n-2}|\mathcal{M}_{1:n-1}^{qs})},
\end{align}
which satisfies
\begin{align}
    R'(x_{n-1:1},\ldots,l'_{n-2:1})=R'(x_{n-1},x_1,\ldots,l'_{n-2:1})\notag \\
    =\ln \frac{(\Pi^{x_1}|\rho_0)}{(\Pi^{x_{n-1}}|\rho_{n-1})}+\sum_{m=1}^{n-2}\ln(Z^{{\gamma}_m^{-1}}_{i_mj_m}Z^{\mathcal{N}_m(\gamma_m)}_{k'_ml'_m}).
\end{align}
We can similarly define a distribution of entropy production $p(R'_1)$ and prove the detailed FT. However, the obtained fluctuation relation is related to processes (\ref{APR1}), not to processes (\ref{NSMBP}). The entropy production $R_1$ and $R'_1$ share the same forward processes, but use different initial states in the backward processes. Obviously,
\begin{align}
    R'(x_{n-1:1},\ldots,l'_{n-2:1})+ R(x_{n:n-1},\ldots,l'_{n-1})\notag \\
    = R(x_{n:1},\ldots,l'_{n-1:1}).
\end{align}
Similarly, we can use different initial states in the forward processes 
\begin{equation}\label{DISFP}
    |\mathcal{S}_{n:n-1}):=\mathcal{N}^{S_{n} }_{n-1}|\tilde{\rho}_{1}^{S_{n-1}}\otimes [ \Phi^{A^{(n-1)}S_{n}}]) 
    \end{equation}
to obtain another entropy production $R'_2$, which satisfies
\begin{align}
    R(x_{n-1:1},\ldots,l'_{n-2:1})+ R'(x_{n:n-1},\ldots,l'_{n-1})\notag \\
    = R(x_{n:1},\ldots,l'_{n-1:1}).
\end{align}

Similar to \cref{RBTPSP}, if both $R$ and $R_1$ satisfy the detailed FT, then 
\begin{equation}\label{RBTPSP1}
   \braket{ e^{-[R(x_{n:1},\ldots,l'_{n-1:1})-R(x_{n-1:1},\ldots,l'_{n-2:1})]}}=1.
\end{equation}
With Jensen's inequality, the conclusion that the total average entropy production is not less than intermediate average entropy production still holds.  And the entropy production rate is still nonnegative. Notice that the forward process of $R'_2$ is different from that of $R$ and $R_1$, so we have in general
\begin{align}
    \braket{R}-\braket{R_1}= \braket{R(x_{n:1},\ldots,l'_{n-1:1})-R(x_{n-1:1},\ldots,l'_{n-2:1})}\notag \\
    =\sum\mathcal{P}_{n:1}(x_{n:1},\ldots,l'_{n-1:1}|\mathcal{M}_{n:1}^{qs}) R'(x_{n:n-1},\ldots,l'_{n-1})\notag \\
    \neq\sum\mathcal{P}_{n:n-1}(x_{n:n-1},\ldots,l'_{n-1}|\mathcal{M}_{n:n-1}^{qs}) R'(x_{n:n-1},\ldots)\notag \\
    =:\braket{R'_2}',
\end{align}
where $\mathcal{P}_{n:n-1}$ is the probability distribution from two-point measurement of the processes (\ref{DISFP}) and $\braket{\cdot}'$ is the corresponding probability average. 

Similar to \cref{EPRA}, using \cref{QSNM,MFTQ,MRHON}, the average of entropy production $R$ here can be generally expressed as
\begin{align}\label{MEPRA}
    \braket{R}=S(\mathcal{M}_{n} \rho_{n}||\mathcal{M}_{n}\tilde{\rho}_0)+S(\mathcal{M}_{n} \rho_{n})-S(\mathcal{M}_{1}\rho_0) \notag \\
    +\sum_{m=1}^{n-1}[\Tr (\rho_{m+1}\ln \mathcal{N}_m \gamma_m)-\Tr(\mathcal{M}_{m} \rho_{m}\ln \gamma_m)]\notag\\
    =S(\mathcal{M}_{n} \rho_{n}||\mathcal{M}_{n}\tilde{\rho}_0)+\sum_{m=2}^{n}[S(\mathcal{M}_{m} \rho_{m})-S(\rho_m)]\notag\\
    +\sum_{m=1}^{n-1}[S(\mathcal{M}_{m} \rho_{m}||\gamma_m)-S( \rho_{m+1}||\mathcal{N}_m \gamma_m)].
\end{align}
Comparing \cref{EPRA,MEPRA}, it is easy to see that both of them contain the term $\Delta S_{\mathcal{M}}=S(\mathcal{M} \rho)-S(\rho)$, which is the direct contribution of the measurements to the entropy production \cite{DPZ16}. If we choose $\rho_n=\tilde{\rho}_0$, set the reference states $\gamma_{m+1}=\mathcal{N}_m \gamma_m$ and assume that all measurements are non-invasive, then the average of entropy production $R$ can be simplified to
\begin{equation}\label{EPMKNI}
    \braket{R}=S( \rho_0||\gamma_0)-S(\mathcal{N}_{n-1:1}\rho_0 ||\mathcal{N}_{n-1:1} \gamma_0),
\end{equation}
where the evolution map
\begin{equation}\label{MPDV}
    \mathcal{N}_{n-1:1}= \mathcal{N}_{n-1}\circ \ldots \circ \mathcal{N}_1.
\end{equation}
For single-shot CPTP evolution, the average entropy production is \cite{KK19}
\begin{equation}\label{SSEP}
    \braket{R}=S( \rho_0||\gamma_0)-S( \mathcal{N}\rho_0 ||\mathcal{N} \gamma_0),
\end{equation}
where the reference state $\gamma_0$ can be freely chosen according to the needs, and the Gibbs states are a common choice. Since the Gibbs states are fixed points of many processes, the following average entropy production formula is often used \cite{ZMVS17,ABBA16,PVC18}
\begin{equation}\label{GSIFP}
    \braket{R}=S( \rho_0||\rho^{(\beta)})-S( \rho_\tau ||\rho^{(\beta)}).
\end{equation}
But if Gibbs states are not fixed points, then \cref{GSIFP} is inappropriate, see \cite{SE19} for  related discussion.

If we choose $\rho_n=\tilde{\rho}_0$ and assume that all measurements are non-invasive, then from \cref{MEPRA}, we can get
\begin{equation}
    \braket{R}=\sum_{m=1}^{n-1}[S(\rho_{m}||\gamma_m)-S( \rho_{m+1}||\mathcal{N}_m \gamma_m)].
\end{equation}
When the evolution of the system state can be described through an exact time-convolutionless master equation 
\begin{equation}
    \rho_S(t)=e^{\int_0^t \mathcal{L}(\tau)d\tau} \rho_S(0),
\end{equation}
one can choose the instantaneous fixed point $\gamma_t$ as the reference state. $\gamma_t$ is a null eigenvector of the generator of the dynamics. For continuous evolution and measurement, the total average entropy production can be written as 
\begin{equation}
    \braket{R}=-\int_{0}^t d\tau \frac{d}{ds} {\bigg|}_{s=0} S(\rho_S(\tau+s)||\gamma_\tau). 
\end{equation}
It is just the entropy production used in  Ref. \cite{SE19,CB22}.

The average of entropy production $R_1$ and $R_2$ is
\begin{align}\label{MEPR12A}
    \braket{R_1}&=S(\mathcal{M}_{n-1} \rho_{n-1}||\mathcal{M}_{n-1}\tilde{\rho}_1)\notag\\
    &+\sum_{m=2}^{n-1}[S(\mathcal{M}_{m} \rho_{m})-S(\rho_m)]\notag\\
     &+\sum_{m=1}^{n-2}[S(\mathcal{M}_{m} \rho_{m}||\gamma_m)-S( \rho_{m+1}||\mathcal{N}_m \gamma_m)],\notag\\
     \braket{R_2}&=S(\mathcal{M}_{n} \rho_{n}||\mathcal{M}_{n}\tilde{\rho}_0)+S(\mathcal{M}_{n} \rho_{n})-S(\rho_n)\notag\\
     &+S(\mathcal{M}_{n-1} \rho_{n-1}||\gamma_{n-1})-S( \rho_{n}||\mathcal{N}_{n-1} \gamma_{n-1}).
\end{align}
Combining \cref{MEPRA,MEPR12A}, we have
\begin{equation}
    \braket{R_1}+ \braket{R_2}- \braket{R}=S(\mathcal{M}_{n-1} \rho_{n-1}||\mathcal{M}_{n-1}\tilde{\rho}_1)\geq 0.
\end{equation}
Different from the case of unitary evolution, even if the Kolmogorov consistency condition is satisfied, due to the irreversibility of the evolution $\mathcal{N}$, the sum of the segmental entropy production is generally greater than the total average entropy production.

\subsection{Non-Markovian open quantum system}
In general, the non-Markovian processes is indivisible. Under $n-1$ step evolution, the forward process state can be written as
\begin{equation}\label{NMMTPC}
    |\mathcal{S}_{n:1}):=\mathcal{N}^{S_{n:2} }|\rho_0^{S_1}\otimes [\bigotimes_{j=2}^{n}  \Phi^{A^{(j-1)}S_{j}}]),
\end{equation}
where 
\begin{equation}\label{NMPC}
    \mathcal{N}^{S_{n:2} }=(  I^E|\mathcal{U}^{S_{n} E}_{n-1}\circ \ldots\circ \mathcal{U}^{S_2 E}_1|\rho_0^E).
\end{equation}
With the measurements (\ref{FNPO}), the joint probability for $n$-point measurements can still be expressed as \cref{JPNPM}.
If we use the Petz recovery map of $\mathcal{N}^{S_{n:2} }$ to give backward processes
\begin{equation}
    |\mathcal{S}^{tr}_{1:n}):=\mathcal{R}^{S_{n:2} }_{\gamma_{2:n}}|\tilde{\rho}_0^{S_n}\otimes [\bigotimes_{j=1}^{n-1}  \Phi^{A^{(j-1)}S_{j}}]),
\end{equation}
then the backward process is not linkable \cite{H22}. For such processes, measuring with the measurement (\ref{BNPO}) will lead to an ill-defined probability distribution. One can freshly prepare the system state at each step and obtain the detailed FT for the many-body channels, or insert a special operation and obtain the detailed FT for the derived channel \cite{H22}. The former has no connection with the joint probability distribution $(\mathcal{O}_{n:1}^{(x_n:x_1)}|\mathcal{S}_{n:1})$. The latter has $n-1$ input states, which are independent and there is no time ordered relationship among them. Therefore, both of them are not suitable for studying multi-entropy production under the same processes.

\begin{figure*}[htb]
    \centering
    \includegraphics[width=1\textwidth]{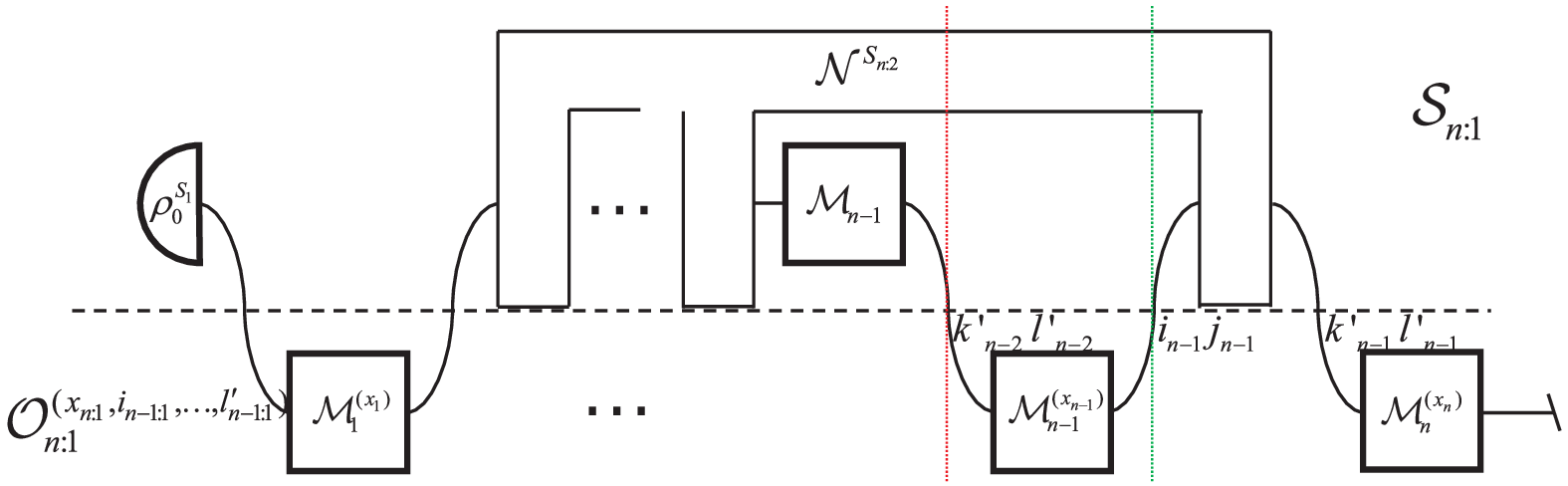}
    \includegraphics[width=1\textwidth]{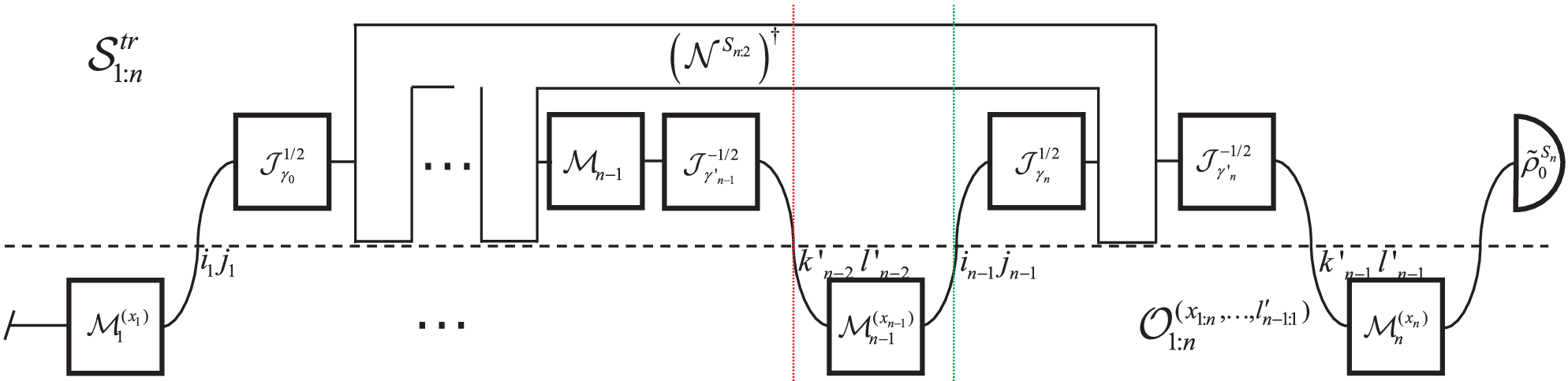}
    \caption{The forward and backward processes of $n-1$ step non-Markovian quantum evolution. The dotted lines represent summing a fraction of the (quasi-)measurements and then obtaining the marginal distributions. Unlike the Markovian cases, the subsequent evolution depends on the state of environment, which in turn depends on the previous measurement results of the system. Therefore, we need  $ \mathcal{N}^{S,E }_{\mathcal{M}}$ to help us find the relation between marginal distributions. The green dotted line corresponds to the division used by $R(x_{n-1:1},\ldots,l'_{n-2:1})$. One needs to sum the measurements to the right of the green dotted line. The red dotted line corresponds to the division used by $ R(x_{n:n-1},\ldots,l'_{n-1})$. One needs to sum the measurements to the left of the red dotted line.}
    \end{figure*}

For non-Markovian processes, the intermediate measurements can influence subsequent evolution of the system and cause entropy increases themselves. Hence, only incorporating the intermediate measurements into the processes itself, can we guarantee that the entropy production is due to the processes rather than the measurements. In this manner, the forward process state is
\begin{equation}
    |\mathcal{S}_{n:1}^{\mathcal{M}}):= \mathcal{N}^{S_{n:2} }_{\mathcal{M}}|\rho_0^{S_1}\otimes [\bigotimes_{j=2}^{n}  \Phi^{A^{(j-1)}S_{j}}]),
\end{equation}
where $ \mathcal{N}^{S_{n:2} }_{\mathcal{M}}=(\bigotimes_{k=2}^{n-1}\mathcal{M}_{k})\circ\mathcal{N}^{S_{n:2} }$. Notice that the induction condition is naturally satisfied in the present framework, the present state and the measurement outcome of the present state cannot be affected by future measurements. Hence, we can obtain the first $m-1$ step evolution by ignoring the results of subsequent evolution $\mathcal{N}^{S_{m:2} }_{\mathcal{M}}=\Tr_{S_{n:m+1}}\mathcal{N}^{S_{n:2}}_{\mathcal{M}}$.
Performing a measurement and discarding the outcomes will not affect the re-measurement of the intermediate state
\begin{equation}
    \mathcal{M}_{k}^{(x_k)}\circ\mathcal{M}_{k}= \mathcal{M}_{k}^{(x_k)}.
\end{equation}
Thus, with the measurements \cref{FNPO}, the joint probability distribution given by $|\mathcal{S}_{n:1}^{\mathcal{M}})$ will be the same as that given by \cref{NMMTPC}. Since the $|\mathcal{S}_{n:1}^{\mathcal{M}})$ has incorporated intermediate measurements into the processes itself, even without further measurements, the final state is also different from that given by \cref{NMMTPC}
\begin{equation}\label{NMEOR}
   |\rho_n)= ([\bigotimes_{i=1}^{n-1}  \Phi^{S_{i}A^{(i)}}]|\mathcal{S}_{n:1}^{\mathcal{M}})\neq  ([\bigotimes_{i=1}^{n-1}  \Phi^{S_{i}A^{(i)}}]|\mathcal{S}_{n:1}).
\end{equation}
For the same reason, their reference final states are also different. 

Here we propose the following backward process state
\begin{equation}
    |\mathcal{S}^{\mathcal{M},tr}_{1:n}):=\mathcal{R}^{S_{n:2}}_{\mathcal{M}}|\tilde{\rho}_0^{S_n}\otimes [\bigotimes_{j=1}^{n-1}  \Phi^{A^{(j-1)}S_{j}}]),
\end{equation}
where 
\begin{equation}\label{NMPRC}
    \mathcal{R}^{S_{n:2}}_{\mathcal{M}}= (\bigotimes_{l=2}^{n}\mathcal{J}_{\gamma_l}^{1/2})\circ(\mathcal{N}^{S_{n:2} }_{\mathcal{M}})^\dagger\circ (\bigotimes_{m=2}^{n}\mathcal{J}_{\gamma'_m}^{-1/2})
\end{equation}
and
\begin{align}\label{IRE}
    \gamma'_m=  ([\bigotimes_{i=1}^{m-1}  \Phi^{S_{i}A^{(i)}}]| (\bigotimes_{k=2}^{\min\{m,n-1\}}\mathcal{M}_{k})\notag \\
    \circ\mathcal{N}^{S_{m:2} }|\gamma_0^{S_1}\otimes [\bigotimes_{j=2}^{m}  \Phi^{A^{(j-1)}S_{j}}]).
\end{align}
$\gamma'_m$ is the output of $\gamma_0$ after $m-1$ steps of evolution. According to this, for $m\leq n-1$, we know that $\gamma'_m$ is the output state of the measurement $\mathcal{M}_{m}$, so they share the same basis. We set $\gamma_2=\gamma_0$ and $\gamma_{l}=\gamma'_{l-1}$ for the others $l$. We still use the quasi-measurements \cref{POVMNM} for forward processes and \cref{BKPOVMNM} for backward processes. Now the basis $\ket{i_m}$ is chosen such that it diagonalizes $\gamma_m$ and $\ket{k'_m}$ is chosen as the eigenbasis of $\gamma'_m$. Relation (\ref{POVMTOPVMMP}) still holds and so does the one of backward processes. 

Before further discussion of the entropy production, we briefly analysis the probability distribution of backward processes. Since $\gamma'_m$ and $\mathcal{M}_{m}$ share the same basis, we have
\begin{equation}
    \mathcal{J}_{\gamma_{m+1}}^{1/2}\circ \mathcal{M}_{m}^{(x_m)}\circ \mathcal{J}_{\gamma'_{m}}^{-1/2} \circ \mathcal{M}_m=\mathcal{M}_{m}^{(x_m)}= \mathcal{M}_{m}^{(x_m)}\circ \mathcal{M}_m.
\end{equation}
Combining this with the definition of $\mathcal{P}^{tr}_{1:n}(x_{1:n}|\mathcal{M}_{1:n}^{qs})$, we obtain
\begin{align}
    \sum_{x_1} \mathcal{P}^{tr}_{1:n}(x_{1:n}|\mathcal{M}_{1:n}^{qs}) = \sum_{x_1} (\mathcal{O}_{1:n}^{(x_{n:1})}|\mathcal{S}^{\mathcal{M},tr}_{1:n})^*\notag\\
     =(\gamma'_{n,x_{2:n-1}}|\mathcal{J}_{\gamma'_n}^{-1/2}\circ\mathcal{M}_{n}^{(x_n)}|\tilde{\rho}_0)^*,
\end{align}
where
\begin{align}
   \gamma'_{n,x_{2:n-1}}=  ([\bigotimes_{i=1}^{n-1}  \Phi^{S_{i}A^{(i)}}]|(\bigotimes_{m=2}^{m-1}\mathcal{M}_{m}^{(x_m)})\circ(\bigotimes_{k=2}^{n-1}\mathcal{M}_{k})\notag \\
    \circ\mathcal{N}^{S_{n:2} }|\gamma_0^{S_1}\otimes [\bigotimes_{j=2}^{n}  \Phi^{A^{(j-1)}S_{j}}])
\end{align}
is related to the intermediate measurement results.
It is easy to show that $\sum_{x_{2:n-1}} \gamma'_{n,x_{2:n-1}}= \gamma'_{n}$, which leads to
\begin{align}
    \sum_{x_{1:n-1}} \mathcal{P}^{tr}_{1:n}(x_{1:n}|\mathcal{M}_{1:n}^{qs}) \notag \\
    =( \gamma'_{n}|\mathcal{J}_{\gamma'_n}^{-1/2}\circ\mathcal{M}_{n}^{(x_n)}|\tilde{\rho}_0)^*=(\Pi_{x_n}|\tilde{\rho}_0).
\end{align}
So the joint probability distribution $\mathcal{P}^{tr}_{1:n}(x_{1:n}|\mathcal{M}_{1:n}^{qs})$ is normalized. 

Now we continue to discuss the FTs. The quasiprobability, entropy production and its probability distribution can be defined similarly to the
\cref{OSMFT}. The entropy production reads
\begin{equation}
    R(x_{n:1},\ldots,l'_{n-1:1})
    =\ln \frac{(\Pi^{x_1}|\rho_0)}{(\Pi^{x_n}|\tilde{\rho}_0)}+\sum_{m=1}^{n-1}\ln(Z^{{\gamma}_{m+1}^{-1}}_{i_mj_m}Z^{\gamma'_{m+1}}_{k'_ml'_m}).
\end{equation}
If we choose $\tilde{\rho}_0=\rho_n$, the average of the entropy production is
\begin{equation}\label{NMFTQAVG}
    \braket{R(x_{n:1},\ldots,l'_{n-1:1})}=S(\rho_0||\gamma_0)-S(\rho_n||\gamma'_n).
\end{equation}
Since the intermediate measurements are already included in the processes $\mathcal{N}^{S_{n:2} }_{\mathcal{M}}$, the applied intermediate measurements are non-invasive. Furthermore, the intermediate reference states cannot be chosen arbitrarily. These make the entropy production (\ref{NMFTQAVG}) quite similar to the form (\ref{EPMKNI}). But note that their evolutions are completely different. The corresponding evolution (\ref{NMEOR}) in entropy production (\ref{NMFTQAVG}) contains measurements and cannot be divided due to memory effects, which are quite different from the evolution (\ref{MPDV}).

Before considering the marginal distribution $\mathcal{P}_{n:1}(x_{n-1:1},i_{n-2:1},\ldots,l'_{n-2:1}|\mathcal{M}_{n:1}^{qs})$, we need to clarify the relation between $ \mathcal{N}^{S_{n-1:2}}_{\mathcal{M}}$ and $ \mathcal{N}^{S_{n:2}}_{\mathcal{M}}$. Unlike the cases in \cref{OSMFT}, the memory effect will make the subsequent evolution depends on the previous measurement results, so the relation is more complex here. From the definition (\ref{NMPC}), we can utilize
\begin{equation}
    \mathcal{N}^{S_{n-1:2},E }_{\mathcal{M}}=(\bigotimes_{k=2}^{n-1}\mathcal{M}_{k})\circ\mathcal{U}^{S_{n-1} E}_{n-2}\circ \ldots\circ \mathcal{U}^{S_2 E}_1
\end{equation}
to bridge $ \mathcal{N}^{S_{n-1:2}}_{\mathcal{M}}$ and $ \mathcal{N}^{S_{n:2}}_{\mathcal{M}}$. Since the measurements on the system will destroy quantum correlations between system and environment \cite{MD11}, we have
\begin{align}\label{RBNSENS}
    (\bigotimes_{m=1}^{n-2} \Pi_{k'_ml'_m}^{S_{m+1}}|\mathcal{N}^{S_{n-1:2},E }_{\mathcal{M}}|[\bigotimes_{m=1}^{n-2} \Pi_{i_mj_m}^{S_{m+1}}]\otimes \rho_0^E)\notag =\\
    (\bigotimes_{m=1}^{n-2} \delta_{k'_ml'_m}  \Pi_{k'_m}^{S_{m+1}}|\mathcal{N}^{S_{n-1:2} }_{\mathcal{M}}|\bigotimes_{m=1}^{n-2} \Pi_{i_mj_m}^{S_{m+1}})\notag\\
    \times|\sigma^E_{i_{n-2:1},j_{n-2:1},k'_{n-2:1}}).
\end{align}
$\sigma^E$ represents the state of the environment.
Since both $\mathcal{M}_{k}$ and $\mathcal{U}$ are CPTP maps, $\mathcal{N}^{S_{n-1:2},E }_{\mathcal{M}}$ is also a CPTP map and this requires
\begin{equation}\label{TPC}
   (I^E |\sigma^E_{i_{n-2:1},j_{n-2:1},k'_{n-2:1}})=\bigotimes_{m=1}^{n-2} \delta_{i_mj_m}. 
\end{equation}
\cref{RBNSENS} can relate $ \mathcal{R}^{S_{n-1:2}}_{\mathcal{M}}$ to $ \mathcal{R}^{S_{n:2}}_{\mathcal{M}}$. According to the definition (\ref{NMPRC}), the recovery map can be rewritten as  
\begin{align}
    \mathcal{R}^{S_{n:2}}_{\mathcal{M}}=(\rho_0^E| (\bigotimes_{l=2}^{n}\mathcal{J}_{\gamma_l}^{1/2})\circ(\mathcal{N}^{S_{n-1:2,E} }_{\mathcal{M}})^\dagger  \notag\\
    \circ(\mathcal{U}^{S_{n} E}_{n-1})^\dagger\circ (\bigotimes_{m=2}^{n}\mathcal{J}_{\gamma'_m}^{-1/2})|I^E).
\end{align}
Combining this with \cref{RBNSENS}, we obtain
\begin{align}\label{RBRS}
    (\bigotimes_{m=1}^{n-1} \Pi_{i_mj_m}^{S_{m+1}}|\mathcal{R}^{S_{n:2}}_{\mathcal{M}}|\bigotimes_{m=1}^{n-1} \Pi_{k'_ml'_m}^{S_{m+1}})= ( \Pi_{i_{n-1}j_{n-1}}^{S_{n}}|\mathcal{J}_{\gamma_n}^{1/2}\notag \\
     \circ(\sigma^E_{i_{n-2:1},j_{n-2:1},k'_{n-2:1}}|(\mathcal{U}^{S_{n} E}_{n-1})^\dagger|I^E)\circ\mathcal{J}_{\gamma'_n}^{-1/2}| \Pi_{k'_{n-1}l'_{n-1}}^{S_{n}})\notag\\
    \times(\bigotimes_{m=1}^{n-2} \Pi_{i_mj_m}^{S_{m+1}}|\mathcal{R}^{S_{n-1:2}}_{\mathcal{M}}|\bigotimes_{m=1}^{n-2} \Pi_{k'_ml'_m}^{S_{m+1}}).
\end{align}
The quasiprobability, entropy production and its probability distribution can also be defined similarly to the \cref{OSMFT}. But unlike \cref{MFTQ1}, according to \cref{RBRS}, the entropy production becomes
\begin{align}\label{MEP1FNMP}
    R(x_{n-1:1},\ldots,l'_{n-2:1})=R(x_{n-1},x_1,\ldots,l'_{n-2:1})\notag \\
    =\ln \frac{(\Pi^{x_1}|\rho_0)}{(\Pi^{x_{n-1}}|\tilde{\rho}^{i_{n-2:1},k'_{n-2:1}}_1)}+\sum_{m=1}^{n-2}\ln(Z^{{\gamma}_{m+1}^{-1}}_{i_mj_m}Z^{\gamma'_{m+1}}_{k'_ml'_m}).
\end{align}
The density matrix will depend on the historical measurements 
\begin{equation}
    \tilde{\rho}^{i_{n-2:1},k'_{n-2:1}}_1=\mathcal{J}_{\gamma_n}^{1/2}\notag \\
    \circ \mathcal{N}_{i_{n-2:1},k'_{n-2:1}}^\dagger\circ\mathcal{J}_{\gamma'_n}^{-1/2}\circ \mathcal{M}_{n}(\tilde{\rho}_0),
\end{equation}
where the evolution depends on the state of the environment
\begin{equation}
    \mathcal{N}_{i_{n-2:1},k'_{n-2:1}}=(I^E|\mathcal{U}^{S_{n} E}_{n-1}|\sigma^E_{i_{n-2:1},i_{n-2:1},k'_{n-2:1}}).
\end{equation}
The detailed FT relation is still held, but the entropy production \cref{MEP1FNMP} is no longer a combination of time-localized measurements, but contains temporal nonlocal measurements. Only when the evolution $ \mathcal{N}_{i_{n-2:1},k'_{n-2:1}}$ does not change with historical measurements, such as the Markovian processes, then the entropy production will only be related to the two-point measurement. 

Now let's discuss another marginal distribution. Summing over $\{x_{n-2:1},i_{n-2:1},\ldots,l'_{n-2:1}\}$ is related to the following processes
\begin{equation}
    \mathcal{N}^{S_{n-1},E }_{\mathcal{M}}=\mathcal{M}_{n-1}\circ\mathcal{U}^{S E}_{n-2}\circ \ldots\circ \mathcal{M}_{2}\circ\mathcal{U}^{SE}_1
\end{equation}
Similar to \cref{RBNSENS}, the measurement makes
\begin{align}
    (\Pi_{k'_{n-2}l'_{n-2}}|\mathcal{N}^{S_{n-1},E }_{\mathcal{M}}|\Pi_{i_1j_1}\otimes \rho_0^E)=|\sigma_{i_1,j_1,k'_{n-2}}^E) \notag\\
    \times \delta_{k'_{n-2}l'_{n-2}} (\Pi_{k'_{n-2}}|\mathcal{N}^{S_{n-1} }_{\mathcal{M}}|\Pi_{i_1j_1}),
\end{align}
where $\mathcal{N}^{S_{n-1} }_{\mathcal{M}}=(I^E|\mathcal{N}^{S_{n-1},E }_{\mathcal{M}}|\rho_0^E)$. With the same reason as \cref{TPC}, it requires $ (I^E |\sigma_{i_1,j_1,k'_{n-2}}^E)= \delta_{i_1j_1} $. With $\mathcal{N}^{S_{n-1},E }_{\mathcal{M}}$, the quasiprobability distribution of forward processes can be written as 
\begin{widetext}
\begin{align}\label{RBNS1}
    \mathcal{P}_{n:1}(x_{n:n-1},\ldots,l'_{n-1}|\mathcal{M}_{n:1}^{qs})\propto(I^E\otimes\Pi_{k'_{n-1}l'_{n-1}}|\mathcal{U}^{S E}_{n-1}|\Pi_{i_{n-1}j_{n-1}}) ( \Pi^{x_{n-1}}|\mathcal{N}^{S_{n-1},E }_{\mathcal{M}}\circ  \mathcal{M}_1|\rho_0\otimes \rho_0^E)
   \notag \\
     =\sum_{i_1,j_1,k'_{n-2}} (\Pi_{k'_{n-1}l'_{n-1}}|\mathcal{N}^{S_n}_{i_1,k'_{n-2}}|\Pi_{i_{n-1}j_{n-1}})\times  (\Pi^{x_{n-1}}|\Pi_{k'_{n-2}})(\Pi_{k'_{n-2}}|\mathcal{N}^{S_{n-1} }_{\mathcal{M}}|\Pi_{i_1j_1})(\Pi_{i_1j_1}|\mathcal{M}_1(\rho_0)),
\end{align}
where $\mathcal{N}^{S_n}_{i_1,k'_{n-2}}=(I^E|\mathcal{U}^{S E}_{n-1}|\sigma_{i_1,i_1,k'_{n-2}}^E)$, and the $\propto$ notation is used because we ignore some common terms like $(\Pi^x|\Pi_{ij})$, which also exist in the quasiprobability distribution of backward processes. The quasiprobability distribution of backward processes is
\begin{align}\label{RBNS2}
    \mathcal{P}^{tr}_{1:n}(x_{n-1:n},\ldots,l'_{n-1}|\mathcal{M}_{1:n}^{qs})\propto\notag \\
   [( I\otimes \rho_0^E|J_{\gamma_{2}}^{1/2} \circ (\mathcal{N}^{S_{n-1},E}_{\mathcal{M}})^\dagger \circ J_{\gamma'_{n-1}}^{-1/2}|\Pi^{x_{n-1}})( \Pi_{i_{n-1}j_{n-1}}| \mathcal{J}_{\gamma_n}^{1/2}\circ(\mathcal{U}^{S E}_{n-1})^\dagger\circ\mathcal{J}_{\gamma'_n}^{-1/2}|\Pi_{k'_{n-1}l'_{n-1}})( \Pi^{x_{n}}|\tilde{\rho}_0\otimes I^E)]^*=\sum_{i_1,j_1,k'_{n-2}} \notag \\
    [(\gamma_0|\Pi_{i_1j_1})(\Pi_{i_1j_1}|(\mathcal{N}^{S_{n-1} }_{\mathcal{M}})^\dagger\circ J_{\gamma'_{n-1}}^{-1/2} |\Pi_{k'_{n-2}}) (\Pi_{k'_{n-2}}|\Pi^{x_{n-1}}) ( \Pi_{i_{n-1}j_{n-1}}|\mathcal{J}_{\gamma_n}^{1/2}\circ(\mathcal{N}^{S_n}_{i_1,k'_{n-2}})^\dagger\circ\mathcal{J}_{\gamma'_n}^{-1/2} |\Pi_{k'_{n-1}l'_{n-1}})( \Pi^{x_{n}}|\tilde{\rho}_0)]^*.
  \end{align}
\end{widetext}
From \cref{RBNS1,RBNS2}, we find that the summing over $\{i_1,j_1\}$ cannot be eliminated or be attributed to a local measurement. So there is no detailed FT for $R(x_{n:n-1},\ldots,l'_{n-1})$ in general. Only when $\mathcal{N}^{S_n}_{i_1,k'_{n-2}}$ is independent of $\{i_1\}$, which is also satisfied in the Markovian processes, then we have
\begin{align}
    R(x_{n:n-1},\ldots,l'_{n-1})=\ln \frac{(\Pi^{x_{n-1}}|\rho_{n-1})}{(\Pi^{x_n}|\tilde{\rho}_0)}\notag \\
    +\ln(Z^{{\gamma}_{n}^{-1}}_{i_{n-1}j_{n-1}}Z^{\gamma'_{n}}_{k'_{n-1}l'_{n-1}}),
\end{align}
where
\begin{equation}
    |\rho_{n-1})=\mathcal{N}^{S_{n-1} }_{\mathcal{M}}\circ\mathcal{M}_1(\rho_0).
\end{equation}
The proof has used that $\gamma'_m$ and $\mathcal{M}_{m}$ share the same basis. From the previous analysis, we can see that if we want the entropy production of the marginal distribution satisfies the fluctuation theorem with time-localized measurements, we must require that the evolution do not change with historical measurements. That is to say, the measurement is not invasive to evolution.

Let's briefly discuss why the marginal distribution in the non-Markovian processes does not necessarily give a detailed FT. Since the proof (\ref{RBTPSP}) also applies to here. If both $R$ and $R_{sub}$ satisfy the detailed FT, then we have $\braket{R}\geq \braket{R_{sub}}$. For non-Markovian processes, the state of the system could be fully recovered. Then according to \cref{NMFTQAVG}, there must be $\braket{R}=0$. However, $\braket{R_{sub}}> 0$ is very natural when there is a contraction in the state space of the system. So there is a conflict here, which makes the detailed FT for marginal distributions not unconditional. 

Since the proof (\ref{RBTPSP}) applies to all cases here, it leads to the following conclusion: For the marginal distribution, if its corresponding entropy production satisfies the detailed fluctuation theorem, then its average gives a lower bound on the total average entropy production.
\section{Example}\label{EX}
To illustrate our framework we briefly discuss the Jaynes-Cummings model. For simplicity, here we consider a two-state atom coupled to
a single harmonic oscillator mode.  Suppose the atom is our system of interest. The non-Markovian dynamics and the memory effects of this model have been discussed by \cite{SVN10,DL13}. Ref. \cite{CB22} took it as an example to calculate the entropy production of a single-shot evolution under different evolution durations, so as to obtain the entropy production rate. The total Hamiltonian is given by
\begin{equation}
    H_{\text{JC}}=\omega_a \frac{\sigma_z}{2}+\omega_c a^\dagger a+\frac{\Omega}{2}(a\sigma_++a^\dagger\sigma_-).
\end{equation}
Its eigenstate is
\begin{align}
    \ket{n,+}&=\cos (\alpha_n/2)\ket{n,1}+\sin  (\alpha_n/2)\ket{n+1,0}\notag\\
    \ket{n,-}&=\sin  (\alpha_n/2)\ket{n,1}-  \cos(\alpha_n/2)\ket{n+1,0},
\end{align}
where $n$ denotes the number of radiation quanta in the mode, the letters $1,0$ denote the excited and ground state respectively,  $\alpha_n=\tan^{-1}(\Omega\sqrt{n+1}/\Delta) $ and $\Delta=\omega_a-\omega_c$.  The energy eigenvalues associated with the eigenstates $ \ket{n,\pm}$  are given by 
\begin{equation}
    E_{n,\pm}=\omega_c(n+\frac{1}{2})\pm \frac{1}{2}\Omega_n,
\end{equation}
where $\Omega_n=\sqrt{\Delta^2+\Omega^2(n+1)}$. The unitary evolution operator $U_{ t}=e^{-iH_{\text{JC}}t}$ in the basis $\{\ket{0},\ket{1}\}$ is given by the following matrix \cite{SVN10}
\begin{equation}\label{OSU}
    U_{ t}=e^{-i\omega_c \hat{n}t}\left(
    \begin{array}{cc}
     c^\dag_{\hat{n}}(t) &-b^\dag d^\dag_{\hat{n}+1}(t)  \\
     d_{\hat{n}+1}(t)b& c_{\hat{n}+1}(t)
    \end{array}
    \right) ,    
\end{equation}
where the operators
\begin{align}
    c_{\hat{n}}(t)&=e^{-i\omega_c t/2}[\cos(\sqrt{\hat{\varphi}}\frac{t}{2}) -i\Delta \sin(\sqrt{\hat{\varphi}}\frac{t}{2})/\sqrt{\hat{\varphi}}], \notag\\
    d_{\hat{n}}(t)&=-ie^{-i\omega_c t/2}2g \sin(\sqrt{\hat{\varphi}}\frac{t}{2})/\sqrt{\hat{\varphi}}
\end{align}
and $\hat{\varphi}=\Delta^2+4g^2\hat{n}$. All functions related to the particle number operator satisfy the following relations
\begin{equation}\label{BEON}
    b f_{\hat{n}}(t)= f_{\hat{n}+1}(t)b,\quad b^\dag f_{\hat{n}+1}(t)= f_{\hat{n}}(t)b^\dag.
\end{equation}
Using these relations and
\begin{equation}\label{UNI}
    c^\dag_{\hat{n}}(t) c_{\hat{n}}(t)+\hat{n}  d^\dag_{\hat{n}}(t) d_{\hat{n}}(t)=1,
\end{equation}
one can easily verify the unitarity of $U(t)$. In addition, using
\begin{align}\label{URG}
    c^\dag_{\hat{n}}(\tau_2) c^\dag_{\hat{n}}(\tau_1)-\hat{n}  d^\dag_{\hat{n}}(\tau_2) d_{\hat{n}}(\tau_1)e^{i\omega_c \tau_1}=c^\dag_{\hat{n}}(\tau_1+\tau_2)\notag\\
    c_{\hat{n}}(\tau_2) d_{\hat{n}}(\tau_1)+ d_{\hat{n}}(\tau_2) c^\dag_{\hat{n}}(\tau_1)e^{-i\omega_c \tau_1}=d_{\hat{n}}(\tau_1+\tau_2),
\end{align}
it is easy to verify that $U(\tau_2)U(\tau_1)=U(\tau_1+\tau_2)$.

Suppose the initial state of the system and the environment is factorized $\rho_0^{SE}=\rho_0^{S}\otimes\rho_0^{E}$, the environment is initially in thermal equilibrium (Gibbs) state $\rho_0^{E}=\exp(-\beta H_c)/Z_c$. Then the dynamical map of the system is
\begin{equation}
    \mathcal{N}_{0\to t}(\rho_0^{S})=\Tr_E(\mathcal{U}_{t}\rho_0^{S}\otimes\rho_0^{E}).
\end{equation}
Since the unitary evolution obeys the energy conservation relation $[H_0,H_I]=0$, the  thermal equilibrium state of system $\rho_S^{(\beta)}=\exp(-\beta H_a)/Z_a$ is the fixed point of map $ \mathcal{N}_{0\to t}$. For single-shot evolution, the corresponding forward process state is
    \begin{align}\label{OOPPS}
        |\mathcal{S}_{2:1})=\mathcal{N}^{S_2}_{0\to t}|\rho_0^{S_1}\otimes   \Phi^{A^{(1)}S_{2}})\notag\\
        =\sum_{ijk'l'}\Phi_{ij\to k'l'}(t)|\rho_0^{S_1}\otimes \Pi_{ij}^{A^{(1)}}\otimes \Pi^{S_{2}}_{k'l'}),
    \end{align}
    where $\Phi_{ij\to k'l'}(t)=\braket{ \Phi_0(t)}_{ij\to k'l'}$. See \cref{PMT} for its specific matrix form.
    
If considering two-step evolution, the corresponding channel is
\begin{equation}
    \mathcal{N}^{S_2S_3}_{0\to t_1\to t_2}(\rho_0^{S_2}\otimes\rho_{t_1}^{S_3})=\Tr_E(\mathcal{U}^{S_2E}_{ t_2-t_1}\mathcal{U}^{S_1E}_{ t_1}\rho_0^{S_2}\otimes\rho_{t_1}^{S_3}\otimes\rho_0^{E}).
\end{equation}
The corresponding forward process state is given by
\begin{align}
    |\mathcal{S}_{3:1})&=\mathcal{N}^{S_2S_3}_{0\to t_1\to t_2}|\rho_0^{S_1}\otimes   \Phi^{A^{(1)}S_{2}}\otimes   \Phi^{A^{(2)}S_{3}})\notag\\
    &=\sum_{i_{2:1}j_{2:1} k'_{2:1}l'_{2:1}}\Phi_{i_{2:1}j_{2:1}\to k'_{2:1}l'_{2:1}}(t_1,t_2) \notag\\
   &\times|\rho_0^{S_1}\otimes \Pi_{i_1j_1}^{A^{(1)}}\otimes \Pi^{S_{2}}_{k_1l_1}\otimes \Pi_{i_2j_2}^{A^{(2)}}\otimes \Pi^{S_{3}}_{k_2l_2}).
\end{align}
The specific matrix form of $\Phi_{i_{2:1}j_{2:1}\to k'_{2:1}l'_{2:1}}(t)$ can be found in \cref{CFOTPM}.

\begin{figure}[htb]
    \centering
    \includegraphics[width=0.48\textwidth]{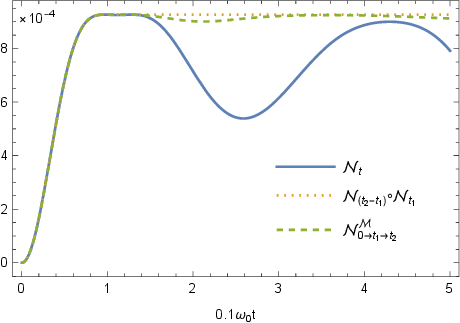}
    \caption{The average entropy production for the Jaynes-Cummings model under different processes. $\mathcal{N}_t$ is single-shot evolution.  $\mathcal{N}_{t_2-t_1}\circ\mathcal{N}_{t_1}$ is two-step Markovian evolution. We applied a measurement to the system at time $\omega_0t_1=10$ while simultaneously refresh the environment to the initial thermal state. $\mathcal{N}^{\mathcal{M}}_{0\to t_1\to t_2}$ is two-step non-Markovian evolution. We applied a measurement to the system at time $\omega_0t_1=10$, but do no additional operations on the environment.  Parameters: $\omega_a=\omega_0$, $g=0.1\omega_0$, $\omega_c=0.1\omega_0$, $k_{\text{B}}T=\omega_0$, $\rho_S^{11}(0)=0.25$ and $\rho_S^{01}(0)=0$. All the measurements are in the basis $\{\ket{0},\ket{1}\}$.
    }\label{AEPCDP}
    \end{figure}

Here, we plot the comparison between the entropy production (\ref{EPMKNI}), (\ref{SSEP}) and (\ref{NMFTQAVG}) (see \cref{AEPCDP}). Comparing the entropy production of $\mathcal{N}_t$ with $\mathcal{N}_{t_2-t_1}\circ\mathcal{N}_{t_1}$, we can find that the memory effect reduces entropy production. When the memory is erased (by refreshing the environment state), the restoration of the state will be prevented. Comparing $\mathcal{N}_t$ with $\mathcal{N}^{\mathcal{M}}_{0\to t_1\to t_2}$, it can be found that the recovery of the state is weakened, which means that the measurement can destroy part of the memory. This is consistent with the fact that memory can be divided into quantum and classical parts.  Comparing $\mathcal{N}_{t_2-t_1}\circ\mathcal{N}_{t_1}$ with $\mathcal{N}^{\mathcal{M}}_{0\to t_1\to t_2}$, it can be found that the negative entropy production still exists, which means that the measurement will not destroy all the memories. In addition, for process $\mathcal{N}^{\mathcal{M}}_{0\to t_1\to t_2}$, we can see that $\braket{R_{\omega_0t=10}}>\braket{R_{\omega_0t=20}} $. This makes it impossible for the corresponding marginal distribution to give a detailed FT.

\section{Conclusion and discussion}\label{CD}
In this paper, we have studied the entropy production and the detailed FT for multitime quantum processes in the framework of the operator-state formalism. For closed quantum systems and Markovian open quantum systems, the entropy production of the joint probability and marginal distributions all satisfy the detailed FT relation. This also leads to a non-negative entropy production rate. And the total average entropy production is not less than intermediate average entropy production. For closed quantum systems, we further showed that the sum of each intermediate amount entropy production can be equal to the total entropy production if the Kolmogorov condition is satisfied. For non-Markovian open quantum systems, the memory effect can lead to a negative entropy production rate. Therefore, the entropy production of the marginal distribution generally does not satisfy the detailed FT relation.

To illustrate the framework, we have briefly calculated the total average entropy production of the three processes in the Jaynes-Cummings model. The results show that memory effects do reduce the average entropy production, while measurements destroy part of the memory. For the non-Markovian open quantum system, the intermediate measurements can destroy the system-environment quantum correlations and thus part of the memory. If these intermediate measurements are not performed, then the full memory effect can be preserved. If the environment state is refreshed at each step, as in \cite{MRA12}, then the evolution is completely Markovian, and the memory effect is completely destroyed. A further discussion of the effect of measurement on different memories, and the effect of different memories on entropy production will help us to fully understand the influence of memory effects on the fluctuation theorems.

This paper mainly focuses on the fluctuation theorem of multiple entropy production under the same process and the same measurement. The multiple fluctuations here are obtained by calculating the marginal distribution of measurements at different times. Some previous studies such as \cite{ZWZW22} have considered the fluctuation theorem under the single-shot process and the multi-body measurement, where the multiple fluctuations are obtained by calculating the marginal distribution of different measurement objects (such as system and reference). The conditional mutual information obtained by introducing the auxiliary system can be used to measure the memory effect. Therefore, studying the fluctuation theorem of many-body measurements under multi-time evolution can give some interesting results.

In \cite{MEJ19,HTNM19}, the entropy production and the detailed FT relation of multiple channels are used to develop an arrow of time statistic associated with the measurement dynamics. Evolution with memory effects is clearly beyond these frameworks, so irreversibility may be violated. The framework of this paper allows us to discuss path probabilities of non-Markovian processes, and thus may help us to deepen our understanding of the relation between the Poincar\'e recurrence theorem and the statistical arrow of time.

The FTs considered here are completely general but only useful when $R$ can be expressed exclusively in terms of physical and measurable quantities. For closed quantum systems and Markovian open quantum systems, the entropy production defined here is consistent with previous work. So I won't go into details here. The main difference is in the non-Markovian cases, where memory effects cause the entropy production to include temporally nonlocal measurements. This issue needs further investigation.

\begin{acknowledgments}
    Z.H. is supported by the National Natural Science Foundation of
China under Grants No. 12305035. 
\end{acknowledgments}

\appendix
\section{Overall process state and open system process state}
Here we use the system-environment unitary evolution process state to obtain the open system process state. The overall process state corresponding to a single-shot unitary evolution is
    \begin{align}
        |\mathcal{S}^{U}_{2:1}):=\mathcal{U}^{S_2E}_{t}|\rho_0^{S_1}\otimes   \Phi^{A^{(1)}S_{2}}\otimes \rho_0^E)\notag\\
        =\sum_{ijk'l'}|\rho_0^{S_1}\otimes \Pi_{ij}^{A^{(1)}}\otimes \Pi^{S_{2}}_{k'l'}\otimes \mathcal{T}^{ij, k'l'}_t(\rho_0^E)),
    \end{align}
    where $\mathcal{T}^{ij, k'l'}_t(\rho)$ can be expressed in the following matrix
    \begin{widetext}
    \begin{equation}
        \mathcal{T}_t(\rho)= e^{-i\omega_c \hat{n}t}\left(
\begin{array}{cccc}
 c^\dag_{\hat{n}}(t)\rho c_{\hat{n}}(t) & c^\dag_{\hat{n}}(t)\rho b^\dag d^\dag_{\hat{n}+1}(t) & d_{\hat{n}+1}(t)b\rho c_{\hat{n}}(t) & d_{\hat{n}+1}(t)b \rho b^\dag d^\dag_{\hat{n}+1}(t)\\
 -c^\dag_{\hat{n}}(t)\rho d_{\hat{n}+1}(t)b & c^\dag_{\hat{n}}(t)\rho c^\dag_{\hat{n}+1}(t) & -d_{\hat{n}+1}(t)b\rho d_{\hat{n}+1}(t)b & d_{\hat{n}+1}(t)b\rho c^\dag_{\hat{n}+1}(t) \\
 -b^\dag d^\dag_{\hat{n}+1}(t)\rho c_{\hat{n}}(t) & -b^\dag d^\dag_{\hat{n}+1}(t)\rho b^\dag d^\dag_{\hat{n}+1}(t) & c_{\hat{n}+1}(t)\rho c_{\hat{n}}(t) & c_{\hat{n}+1}(t)\rho b^\dag d^\dag_{\hat{n}+1}(t)\\
 b^\dag d^\dag_{\hat{n}+1}(t)\rho d_{\hat{n}+1}(t)b & -b^\dag d^\dag_{\hat{n}+1}(t)\rho c^\dag_{\hat{n}+1}(t)  & -c_{\hat{n}+1}(t)\rho d_{\hat{n}+1}(t)b & c_{\hat{n}+1}(t)\rho c^\dag_{\hat{n}+1}(t) \\
\end{array}
\right)e^{i\omega_c \hat{n}t}
    \end{equation}
    where the rows are $ij$ and the columns are $k'l'$, and the value is $\{00,01,10,11\}$. According to \cref{NMPC}, we can obtain the process state of the open system from the overall process state. When the environmental state commuting with the number operator, we have  $(I^E|\mathcal{T}_t(\rho_0^E))=\Phi_0(t)$, where
    \begin{equation}\label{PMT}
        \Phi_m(t)= \left(
            \begin{array}{cccc}
             \alpha_m ( t) & 0 & 0 & 1-\alpha_m  (t) \\
             0 & \gamma^\dag_m  (t)& 0 & 0 \\
             0 & 0 & \gamma_m (t) & 0 \\
             1-\alpha_{m+1} (t)  & 0 & 0 & \alpha_{m+1}  (t) \\
            \end{array}
            \right),     
    \end{equation}
    $ \alpha_{i}(t)= c^\dag_{\hat{n}+i}(t) c_{\hat{n}+i}(t)$ and $\gamma_{i}(t)= c_{\hat{n}+i}(t) c_{\hat{n}+i+1}(t)$.
    For two-step unitary evolution, the following overall process state can be obtained
    \begin{align}
        |\mathcal{S}^{U}_{3:1}) =\sum_{i_{2:1}j_{2:1} k'_{2:1}l'_{2:1}}|\rho_0^{S_1}\otimes \Pi_{i_1j_1}^{A^{(1)}}\otimes \Pi^{S_{2}}_{k'_1l'_1}\otimes \Pi_{i_2j_2}^{A^{(2)}}\otimes \Pi^{S_{3}}_{k_2l_2}\otimes \mathcal{T}^{i_2j_2, k'_2l'_2}_{t_2-t_1}[\mathcal{T}^{i_1j_1, k'_1l'_1}_{t_1}(\rho_0^E)]).
    \end{align}
    Again, we can use $(I^E|\mathcal{T}_{t_2-t_1}[\mathcal{T}_{t_1}(\rho_0^E)])$ to obtain the specific form of the map $\Phi_{i_{2:1}j_{2:1}\to k'_{2:1}l'_{2:1}}(t_1,t_2)$. Using \cref{BEON,UNI,URG} repeatedly, one can get
    \begin{align}\label{CFOTPM}
        \Phi_{i_1j_1\to k'_1l'_1,X_1}=\braket{ \Phi_0(t_1)\Phi_{\Delta_1}(t_2-t_1)}_{i_1j_1\to k'_1l'_1,X_1}\notag \\
        \Phi_{i_1j_1\to k'_1l'_1,X_2}=\delta_{\Delta_2+i_1-j_1-k'_1+l'_1,0}\braket{\Phi^{s_1}(t_1,t_2)\Phi^{s_2}_{\Delta_1+1/2}(t_2-t_1)}_{i_1j_1\to k'_1l'_1,X_2} \notag \\
        \Phi_{i_1j_1\to k'_1l'_1,X_3}=\delta_{i_1,i^\bot_2}\delta_{j_1,j^\bot_2}\delta_{k'_1,k'^\bot_2}\delta_{l'_1,l'^\bot_2}\braket{\xi^\dag_0(t_1,t_2)\xi_1(t_1,t_2)}
    \end{align}
   where $X_1=\{i_2j_2\to i_2j_2, i_2i_2\to i^\bot_2i^\bot_2 \}$, $X_2=\{i_2j_2\to i^\bot_2j_2, i_2j_2\to i_2j^\bot_2 \}$, $X_3=\{i_2i^\bot_2 \to i^\bot_2i_2 \}$,  $\Delta_1=(i_1+j_1-k'_1-l'_1)/2$, $\Delta_2=(i_2-j_2-k'_2+l'_2)$ and
    \begin{align}
\Phi^{s_1}(t_1,t_2)&= \left(
\begin{array}{cccc}
 0 & \xi_0(t_1,t_2)c^\dag_{\hat{n}}(t_1)  & \xi^\dag_0(t_1,t_2)c_{\hat{n}}(t_1)  & 0 \\
 -\xi^\dag_1(t_1,t_2)c^\dag_{\hat{n}}(t_1)  & 0 & 0 & \xi^\dag_0(t_1,t_2) c^\dag_{\hat{n}+1}(t_1) \\
 -\xi_1(t_1,t_2)c_{\hat{n}}(t_1) & 0 & 0 & \xi_0(t_1,t_2) c_{\hat{n}+1}(t_1)\\
 0 &  -\xi_1(t_1,t_2)c^\dag_{\hat{n}+1}(t_1) &  -\xi^\dag_1(t_1,t_2)c_{\hat{n}+1}(t_1)  & 0 \\
\end{array}
\right) \\
\Phi^{s_2}_m(t)&= \left(
    \begin{array}{cccc}
     0 & c^\dag_{\hat{n}+m-1}(t) & c_{\hat{n}+m-1}(t) & 0 \\
     -c^\dag_{\hat{n}+m}(t)& 0& 0 & c^\dag_{\hat{n}+m}(t) \\
     -c_{\hat{n}+m}(t) & 0 & 0 & c_{\hat{n}+m}(t) \\
     0  & -c^\dag_{\hat{n}+m+1}(t) & -c_{\hat{n}+m+1}(t) & 0 \\
    \end{array}
    \right),     
    \end{align}
    where 
    \begin{equation}
        \xi_i(t_1,t_2)=c_{\hat{n}+i}(t_1) c_{\hat{n}+i}(t_2-t_1)- c_{\hat{n}+i}(t_2)=e^{-i\omega_ct_2}\xi^\dag_i(t_1,t_2).
    \end{equation}
    When $t_2=t_1$, there is $\xi_i(t_1,t_2)=0$, and it is easy to verify that
    \begin{equation}
        \Phi_{i_1j_1\to k'_1l'_1,i_2j_2\to k'_2l'_2}=\braket{ \Phi_0(t_1)}_{i_1j_1\to k'_1l'_1}\delta_{i_2,k'_2}\delta_{j_2,l'_2}.
    \end{equation}
    This mapping does correspond to the channel $\Tr_E(\mathcal{I}^{S_2E}\circ\mathcal{U}^{S_1E}_{0\to t_1}(\cdot\otimes\rho_0^{E}))$. In addition, one-step evolution can be derived from two-step evolution when intermediate states are directly connected. Accordingly, it can be verified that
    \begin{equation}
        \sum_{k'_2,i_2,l'_2,j_2}\delta_{k'_2,i_2}\delta_{l'_2,j_2}\Phi_{i_{2:1}j_{2:1}\to k'_{2:1}l'_{2:1}}(t_1,t_2)=\Phi_{i_1j_1\to k'_2l'_2}(t_2).
    \end{equation}
    If the intermediate states are measured, and assuming that the measurement is in the basis $\{\ket{0},\ket{1}\}$, then from
    \begin{equation}
        \sum_{k'_2,i_2,l'_2,j_2}\delta_{k'_2,i_2}\delta_{l'_2,j_2}\delta_{k'_2,l'_2}\Phi_{i_{2:1}j_{2:1}\to k'_{2:1}l'_{2:1}}(t_1,t_2)=\Phi^{\mathcal{M}}_{i_1j_1\to k'_2l'_2}(t_1,t_2)
    \end{equation}
    we can get the mapping corresponding to \cref{NMEOR}. Its specific matrix form is
    \begin{equation}\label{NMEORCF}
        \Phi^{\mathcal{M}}(t_1,t_2)= \left(
            \begin{array}{cccc}
             1-\chi_0(t_1,t_2) & 0 & 0 & \chi_0(t_1,t_2) \\
             0 & \eta^\dag(t_1,t_2) & 0 & 0 \\
             0 & 0 & \eta(t_1,t_2) & 0 \\
             \chi_1(t_1,t_2)  & 0 & 0 & 1-\chi_1(t_1,t_2) \\
            \end{array}
            \right)  
    \end{equation}
    where 
    \begin{equation}
        \chi_i(t_1,t_2)=\alpha_i ( t_1)+\alpha_i ( t_2-t_1)-2\alpha_i ( t_1)\alpha_i ( t_2-t_1),  \eta(t_1,t_2)=- \xi_0(t_1,t_2)c_{\hat{n}+1}(t_1) c_{\hat{n}+1}(t_2-t_1)- \xi_1(t_1,t_2)c_{\hat{n}}(t_1) c_{\hat{n}}(t_2-t_1).
    \end{equation}
    The mapping \cref{NMEORCF} can be used to calculate the entropy production \cref{NMFTQAVG}.

 \end{widetext}

\end{CJK*}

\end{document}